\documentclass[12pt,preprint]{aastex}
\newcommand\kms    {\ifmmode{{\rm ~km~s}^{-1} }\else{~km~s$^{-1}$ }\fi}

\shorttitle{Clusters in Dwarf Irregular Galaxies}
\shortauthors{Seth, Olsen \& Miller}

\begin{document}

\slugcomment{Accepted for publication in Feb. 2004 AJ}

\title{Star Clusters in Virgo and Fornax Dwarf Irregular Galaxies}

\author{Anil Seth}
\affil{University of Washington}
\email{seth@astro.washington.edu}

\author{Knut Olsen}
\affil{Cerro Tololo Interamerican Observatory}
\email{kolsen@ctio.noao.edu}

\author{Bryan Miller}
\affil{Gemini South}
\email{bmiller@gemini.edu}

\author{Jennifer Lotz}
\affil{University of California, Santa Cruz}
\email{jlotz@scipp.ucsc.edu}
\and
\author{Rosie Telford}
\affil{University of Leicester}
\email{rosie.telford@astro.le.ac.uk}

\begin{abstract}

We present the results of a search for clusters in dwarf irregular
galaxies in the Virgo and Fornax Cluster using HST WFPC2 snapshot
data.  The galaxy sample includes 28 galaxies, 11 of which are
confirmed members of the Virgo and Fornax clusters.  In the 11
confirmed members, we detect 237 cluster candidates and determine
their V magnitudes, V-I colors and core radii.  After statistical
subtraction of background galaxies and foreground stars, most of the
cluster candidates have V-I colors of -0.2 and 1.4, V magnitudes lying
between 20 and 25th magnitude and core radii between 0 and 6 pc.
Using H$\alpha$ observations, we find that 26\% 
of the blue cluster candidates are most likely HII regions.  The rest
of the cluster candidates are most likely massive ($>$10$^4$ M$_\odot$)
young and old clusters.  A
comparison between the red cluster candidates in our sample and the
Milky Way globular clusters shows that they have similar luminosity
distributions, but that the red cluster candidates typically have
larger core radii.  Assuming that the red cluster candidates are in
fact globular clusters, we derive specific frequencies (S$_N$) ranging
from 
$\sim$0-9 for the galaxies.  Although the values are uncertain, seven of
the galaxies appear to have specific frequencies greater than 2.  These values
are more typical of ellipticals and nucleated dwarf ellipticals than
they are of spirals or Local Group dwarf irregulars.

\end{abstract}

\section{Introduction}

The study of extragalactic globular cluster systems in galaxies of a
variety of Hubble types has led to the development of a number of
scenarios that place globular clusters within the context of galaxy
formation and evolution.  \citet{carney01} and \citet{vandenbergh00b}
reviewed the myriad pieces of evidence derived from globular cluster
systems and the scenarios designed to explain them.  The properties of
globular clusters in giant elliptical galaxies, in particular, have
led to suggestions that globular cluster systems form 1) {\it in
situ}, perhaps in the hierarchical collapse of the halo \citep{cote00} 
and possibly in multiple phases \citep{forbes97}; 2) in galaxy mergers, 
both through the formation of new clusters in the colliding gas 
\citep[e.g.][]{schweizer96} and 
through the combination of the pre-existing globular cluster systems
\citep[e.g.][]{forbes00}; 3) by accretion of the clusters of smaller
galaxies \citep{cote98}, as exemplified by the $\sim$4 GCs
being contributed to the Milky Way \citep{dacosta95} by
the Sagittarius dwarf galaxy \citep{ibata94}.  More recently the
second two of these ideas have been incorporated into the current
cosmological scenario using simulations \citep[e.g.][]{kravtsov03} and
semi-analytic methods \citep[e.g.][]{beasley02}.

The generally low globular cluster specific frequencies 
\citep[$S_N$, the number of clusters normalized by the galaxy magnitude;][]{harris81} observed in nearby dwarf and spiral galaxies
pose a difficulty for the third of the above possibilities, a point
which was raised by \citet{vandenbergh82} in argument against
spiral-spiral mergers explaining the high $S_N$ of cD galaxies.
However, it may be misleading to compare the specific frequences of
spiral and dwarf galaxies residing in nearby low-density galaxy groups
with those of giant elliptical galaxies found in more distant rich
clusters of galaxies, since the role of the environment in the
assembly of globular cluster systems is not understood.  A census of
the globular cluster systems of a wide variety of galaxies drawn from
the same environment is needed to properly assess the roles that each
of the formation mechanisms may play.

\citet{miller98} embarked on such a census through the
study of the globular clusters of dwarf elliptical (dE) galaxies in
the Virgo, Fornax, and Leo clusters.  The results of that survey
revealed a distinction between the globular cluster populations of
nucleated dE (dE,N) and nonnucleated dE galaxies; in particular, dE,N
galaxies were found to have a factor of 3 higher $S_N$ than
nonnucelated dEs.  The high $S_N$ of dE,N align them more closely with
giant elliptical galaxies than with dwarf irregular (dIrr) and spiral
galaxies, which from measurements in {\it local} galaxies have
$S_N\sim0.5$ \citep{harris91}; the nonnucleated dE's, on the other hand,
{\it do} have $S_N$ that resemble those of dIrrs and spiral galaxies,
once fading of the young stellar populations is taken into account.
The difference in $S_N$ between dE and dE,N galaxies thus suggests
globular cluster formation processes in dE,N and giant ellipticals
that are different from those of spirals and dIrrs.  Moreover, it
argues that not all dE galaxies represent the stripped or harassed
remnants of dIrr galaxies, processes described by 
e.g. \citet{lin83} and \citet{moore98}.

These conclusions depend, however, to large degree on the validity of
comparing globular cluster populations in rich-cluster dE galaxies
with those of local dIrr and spiral galaxies.  In this paper, we
discuss the results of a {\it Hubble Space Telescope} (HST) snapshot
survey of dIrr galaxies in the Virgo and Fornax galaxy clusters, the
same environments from which many of the dE galaxies of \citet{miller98} 
were drawn.  We describe the observational sample and data
reduction in \S2.  In \S3, we describe our technique for detecting
candidate globular clusters.  We present the properties of our sample
of candidate globular clusters, derive specific frequencies, and
examine individual galaxies in \S4.  Some of the more
interesting results are discussed in \S5, followed by our conclusions
in \S6.

\section{Observations and the Galaxy Sample}

The galaxy sample, shown in Table~1, was observed as part of a snapshot
program GO-7377 
(PI: Miller) using the WFPC2 camera of HST.  All HST observations
consisted of two 240 second exposures in the F555W filter and one 300
second exposure in the F814W filter.  The galaxies, which have typical
sizes of $\sim$30 arcsec, were in all cases centered and completely
contained (except VCC 1448) on the WF3 chip.  The F555W observations,
with their much more accurate cosmic ray removal, are used for most of
the analysis in the paper.  The F814W observations are only used to
derive the V-I colors.

The HST sample originally included 29 galaxies selected from the Virgo
Cluster Catalog \citep[VCC;][]{binggeli85}, 
Fornax Cluster Catalog \citep[FCC;][]{ferguson89}, and
Dorado Group Catalog \citep[DGC;][]{ferguson90} in a
manner similar to that used to select the dE galaxies for 
\citet{miller98}, \citet{lotz01}, and \citet{stiavelli01}.
Targets classified as Im or dE/Im in the catalogs and that do not have
HST guide stars within 3 arcminutes (so as to limit the chance of
having a highly saturated star in the field) were selected. Since
lower luminosity galaxies are more likely to have fewer clusters, more
galaxies were selected in lower-luminosity bins. Since a minority of
galaxies had measured velocities, radial velocity was not used as a
selection criteria.  In addition, the dI/dE transition object NGC~4286
\citep{sandage91} was included as a special
target because it fit a project goal of investigating the connections
between dE and dI galaxies.  Of these 30 galaxies 29 were observed
successfully during HST Cycle~7.  In this paper we concentrate on
the 28 galaxies in the Virgo and Fornax cluster in this sample.

The initial reduction procedure for the HST data was similar to that
given in \citet{miller97}. 
The two F555W images were combined and cleaned using IDL routines
supplied by Richard White of STScI.  All other processing was done
using scripts written in IRAF\footnote{IRAF is distributed by the
National Optical Astronomy Observatories, which are operated by the
Association of Universities for Research in Astronomy, Inc., under
cooperative agreement with the National Science Foundation.}.  The
cleaned F555W images were then 
used as templates to identify and remove cosmic rays in the F814W
images.  An HST image of one of the sample galaxies, VCC 328, can be seen in
Figure~1a.

H$\alpha$ images were taken through a set of four narrow band
(FWHM=20\AA) filters on the Apache Point Observatory 3.5m
telescope in April, 2003.  Each galaxy was observed through an 'ON'
filter tuned to the velocity of the galaxy and an 'OFF' filter,
separated by at least 1000 km/sec from the galaxy velocity.  The
filters had central wavelengths of 6563, 6585, 6607 and 6629 \AA.
Feige 34 was used as a spectrophotmetric standard \citep{massey88}.
Based on 
the scatter in the photometry of Feige 34, we estimate that absolute
fluxes are only accurate to 50\% due to the presence of some clouds
during the first night of observations.  Observations on
subsequent nights allowed for the bootstrapping of the flux
calibration using the galaxy VCC 1374.  
Bias subtraction 
and flat fielding were
performed using CCDPROC.  For one filter (centered at
6585\AA), the flat fielding left 10\% 
variations in the sky level across the field requiring the use of an
illumination correction image.  This illumination correction image was
derived from seven program images with the sources masked using
SEXtractor.  After the illumination correction the sky was constant
to 2\%.  Following flat-fielding, the sky was subtracted from each
image and cosmic rays 
were removed using the IDL routine la\_cosmic \citep{vandokkum01}.
The ON and OFF 
images were then aligned and scaled using the brightest stars in
each field and then were subtracted.  The resulting H$\alpha$ images
(one example is shown in Figure~1b) have detection limits
corresponding to a flux of $\sim 3 \times 10^{35}$ 
ergs/sec at a distance of 17 Mpc.  To determine accurate WCS
coordinates for the images, we used the IRAF MSCRED task MSCZERO
\citep{valdes97}.  Based on the 
scatter of stars relative to their expected positions within the
field, these coordinates are good to within 1$\arcsec$.   

\section{Identifying Star Cluster Candidates}

This section will detail the process by which we identify and derive
properties of candidate clusters.  Before explaining the process in
detail, we provide a brief overview: 

\noindent Using DAOPHOT II \citep{stetson87}, we derived a
point-spread function (PSF) using HST data of the Large Magellanic
Cloud (LMC) cluster NGC 1868.  We convolved this PSF with King
profiles of varying size to 
create model cluster profiles.  Then for each of the galaxies in our
sample, sources were identified and fit to the model cluster profiles
to determine their magnitudes and sizes.  To obtain a list of candidate
star clusters in each 
galaxy, we assumed that all point-like sources on the WF3 chip between our
completeness limit and an M$_V$ of -11.5 are potential clusters.
We then used the flanking WF2 and WF4 chips on the WFPC2 camera to correct
for the presence of background galaxies and foreground stars in our
list of cluster candidates.

An alternate method for finding
cluster candidates in nearby, late-type, external galaxies with HST
data is given by \citet{dolphin02a}.  The methods are similar in that
both detect magnitudes and sizes for cluster candidates in galaxies at
moderate distance, but differ in the method used to determine those
quantities, and the way in which the cluster samples are defined.

For simplicity, throughout the paper we assume a distance to Virgo and
Fornax of 17 Mpc (DM of 31.15) consistent with the distances given in
\citet{ferrarese96} and \citet{silbermann99}.  Differences in the
distance modulus of a few tenths of a magnitude should not strongly
affect our results.

\subsection{Determining the PSF}

To determine a PSF for the WF2, WF3 and WF4 chips in the F555W filter
we used WFPC2 observations of a field of the LMC cluster NGC 1868 that were
taken on Nov. 12, 1998 -- close in time to the majority of our
observations (see Table~1).  The PSFs were 
characterized in DAOPHOT II using roughly 100 stars distributed across
each chip.  The derived PSFs (one for each chip) are a position dependent
combination of a Moffat function with $\beta$=1.5 and sub-pixel
corrections.  The 
PSF was fit using a 5 pixel (0.5 arcsec) radius and extends over a 7
pixel radius.  We conducted a qualitative test comparing our derived
PSFs to Tiny Tim PSFs \citep{krist95} by subtracting both from
point-source fields 
(using SUBSTAR in 
DAOPHOT II) and found that our derived PSFs left behind much less
residual flux.  We will discuss results on the stability of the
PSF in the next section.

Because we are interested in determining the radii of the star cluster
candidates, we created {\it model cluster profiles} by convolving the
derived PSF with King profiles.  Seventeen different King profiles
were used, all with a tidal radius of 40 pc, a value typical of
Galactic globular 
clusters \citep{harris96}, and a core-radius of 0 (the original PSF),
0.25, 0.5, 0.75, 1, 2, 3, 4, 5, 6, 7, 8, 10, 12, 15 and 20 pc  (where
all physical sizes assume a distance of 17 Mpc).  The model
cluster profiles are always referred to by their physical sizes; note
that 1 WFPC2/WF pixel = 0.1 arcsec = 8.2 pc at 17 Mpc. The convolution
of the PSF with the King profiles was done in 
IDL at 5 times the pixel resolution in order to avoid the effects of 
undersampling.  The model cluster profiles were then converted back
into DAOPHOT's PSF format, with a lookup table for the variation of
the PSF as a function of position in the frame and its sub-pixel
position.

\subsection{Photometry and Radii Determination}

A pipeline consisting of DAOPHOT II and IDL programs was used to
identify candidate clusters. 
Sources were found using the DAOPHOT II FIND routine using a threshold
of 5$\sigma$ and a FWHM of 1.4 pixels.  Lower thresholds were found to
give too many spurious sources.  Changing the FWHM was found not to
strongly affect the number/quality of detected sources - partially
resolved (broader) sources were not thrown out by the routine.  However,
asymmetric sources are rejected by the FIND routine, perhaps resulting
in the elimination of some larger HII regions from our sample.

Aperture photometry was then done using the PHOT routine
with a 5 pixel / 0.5 arcsec radius, which is the radius at which the
WFPC2 standard photometric system is defined \citep{holtzman95}.  For
the profile
photometry, we fit each source in the F555W images
to all 17 of the
model cluster profiles in turn (see \S3.1).  For most sources, the
$\chi$ (the quality-of-fit parameter output by DAOPHOT II) of the profile 
fit varies smoothly with the core radius of the model cluster
profile.
This is shown for a number of sources in VCC 1448 in Figure~2.  Note
that many of the minimum $\chi$ values have values very close to
1, confirming that the PSF and model cluster profiles accurately
match the data. 
The fit with the minimum $\chi$ value was used to derive 
photometry and to determine the core radii of the cluster candidates.
Aperture corrections were measured for each of the model cluster
profiles and ranged between +0.026 for the more compact model cluster
profiles and +0.034 for the model cluster profile with
R$_{core}$=20pc.  Since the majority of our cluster candidates have
R$_{core}<5$pc, we applied an aperture correction of
+0.026 to all data.  
Figure~3 shows the
scatter in profile-fit and aperture photometry.  The scatter shows an
asymmetric distribution with the aperture photometry in general being
brighter, probably due to confusion from background and nearby sources.  
In our final sample (crosses), this scatter is significantly
reduced due by our application of a 90\% completeness cut (see \S3.4) and the
exclusion of two high surface brightness regions in VCC 1374. 

For both aperture and profile-fit photometry the F555W and F814W fluxes
were converted into Johnson V and Kron-Cousins 
I magnitudes using the procedure
given in \citet{holtzman95}. In addition, the charge transfer
efficiency was corrected for using the method in \citet{stetson98} and
the geometric distortion was corrected for by multiplying the images by
the pixel area map.  The V magnitudes of all sources were derived from the
profile-fit photometry (V$_{\rm profile}$); however, we don't derive
F814W profile-fit 
photometry and so aperture magnitudes (for both V \& I) are used for
colors.  The 
median error on the V-I color is 0.15 magnitudes for our full sample
of cluster candidates.  In the color range between V-I of 0.5 and 1.5,
critical for separating young and old clusters, the median V-I error is 0.12
magnitudes.  These errors are derived from the errors in the aperture
magnitude as reported by DAOPHOT II.

Magnitudes for the galaxies as a whole were also derived from the
WFPC2 F555W data.  The magnitudes were calculated in two ways: (1) by
computing the total flux within a circular aperture with a radius two
times the pertrosian radius \citep{petrosian76}, and (2) by summing up
all the light in the WF3 chip after masking out clearly extended
background galaxies and very bright foreground stars.  We take the
mean of these two methods as the 
galaxy magnitude, V$_T$, shown in Table~2.
Also shown is the difference in the two magnitudes,
$\Delta$V$_T$.  This number, which is larger than the formal error in
the magnitude calculation, gives a sense of the uncertainty in our
magnitudes.  For brighter galaxies this difference is typically less than
0.1 magnitudes, while it is over 1 magnitude for the faintest
galaxies in our sample.  No color or CTE corrections were
made to the magnitudes, these effects contribute to V$_T$
below the 0.1 magnitude level.  As noted earlier, VCC 1448 is larger
than the WF3 chip field of view, and FCC 76 and VCC 888 also may
contain some emission off the chip; for these galaxies, the magnitudes
given are lower limits.  

As noted above, a measure of the core radius of each candidate cluster
is obtained using the best fitting model cluster profile.  At our
assumed distance of 17 Mpc, one WF pixel corresponds to 8.2 pc.  We
expect to be able to determine core radii smaller than this, 
because the broadening due to the King profiles extends out past the
central pixel; for example, the fraction of light falling outside of
the central pixel changes from 63\% for the PSF to
79\% for the model cluster profile with a core radius of 2 pc. 
Nevertheless, it is important to
consider the effects of our limited spatial resolution and of
the variability in the $HST$ PSF on our measurements of the core radii of
cluster candidates.  We address the issue of the constancy of the PSF
here, and address the effects of resolution and random error
in the next section.

We conducted a test to check whether or not our routine accurately
detected point-sources in seven WFPC2 fields taken during the period
Aug. 2, 1997 to 
Apr. 3, 1999, spanning the dates of our observations.  The fields were
drawn from HST program PIDs 6364, 6804, 7307, 7335, and 7434.  All
fields were either Galactic globular clusters or located in the Large
Magellanic 
Cloud and thus should be dominated by point sources.  We ran them
through the same pipeline as we used for our program sources.   Including
all seven fields, $\sim$80\% of the isolated sources were best fit by
model cluster profiles with core radii less than or equal to 1 pc
(0.012 arcsec).  As a comparison, only $\sim$40\% of
isolated sources in our program sources were best fit by model cluster
profiles with core radii $\le$ 1 pc.
This test shows that our method reliably assigns point sources
small values for the core radius.  It also
allows us to examine the variation of the PSF over time, which would
result in a systematic error in our determinations of the cluster radii.  
For the seven fields, the peak in the best-fitting core radius
distribution varied between 0 and 1 pc.  This suggests that the PSF is
unstable at a level corresponding to $\sim$1 pc (0.012 arcsec) in our
method.  As we will see in the next section, this is roughly the same
level as our random error on a single measurement of the core radius.  

\subsection{Artificial Star/Cluster Tests}

We conducted artificial star and cluster tests using the DAOPHOT II ADDSTAR
routine.  These tests were done by adding grids of objects generated
from the original PSF (R$_{\rm core}$=0 pc) and model cluster profiles
with R$_{\rm core}$ of 1, 3, 7 and 12 pc.  The grids were added
to the WF3 chip galaxy images with a random position offset and
typically contained 600 objects each separated by 30 pixels. Input
magnitudes for the stars were determined randomly and ranged from 
18th to 27th magnitude.  The fields with the added stars/clusters were
then run through the same pipeline outlined in the previous section to
identify and determine properties of the objects.
Because identification and determination of 
the radius was done using the F555W observations, the tests were only done
for those images.  For each galaxy, a minimum of 50,000 artificial objects
were inserted for each of the five model cluster profiles used.  The
artificial star and cluster tests were run using the Condor
distributed computing software\footnote{The Condor Software Program
(Condor) was developed by the Condor Team at the Computer Sciences
Department of the University of Wisconsin-Madison.  All rights, title,
and interest in Condor are owned by the Condor Team.}.  

Curves of  fifty and ninety percent completeness as functions of
R$_{\rm core}$ and F555W magnitude are shown in 
Figure~4a. The completeness is best for point sources
and gets worse with increasing core radius due to the broader profiles
and lower surface brightnesses of these objects. This result is
generally applicable to all surveys of extragalactic globular
clusters -- the completeness limit for point sources can differ by more than
a magnitude from the completeness limit of more extended objects.  This
makes correcting for this completeness function complicated if
there are large numbers of extended sources.  We note that the F555W
magnitudes differ from the Johnson V magnitudes by a color correction
term which is typically small (few $\times 10^{-2}$ magnitudes), and
therefore can be ignored for the completeness limits.  The derived error on
the magnitudes is typically $<$0.01 on the bright end and $\sim$0.2
near the 50\% completeness limit.  Our actual errors, especially on
the bright end, will be somewhat worse than this due to the effects of
pixelization.  

The artificial cluster tests can also be used to determine the random error
in our determination of the core radii of the cluster candidates.
Figure~4b shows the standard deviation of the difference between the
input core radius and the detected core radius ($\sigma$(R$_{\rm core}$)).  For point sources, this number is typically
less than 1 pc, consistent with the results from our point source
fields test.  Most of the galaxies have similar errors at each core
radius, but the brightest galaxies have larger errors caused by the
increased background against which the profiles were fitted.  The
error in core radius also is a function of magnitude - the median
value of $\sigma$(R$_{\rm core}$) is shown for each input core radius
in Figure~5.  The vertical lines indicate the median 90\% completeness
for each input core radius.  

Given the systematic error of up to 1 pc from the point source fields
test, and the random errors shown in Figure~4b and Figure~5, we believe our
measurements of core radius have an error of $\sim$2 pc for objects
brighter than the 90\% completeness limit.  In addition,
the decreased random error for small objects means that objects 
with a best-fitting core radius of $\ge$2 pc can reliably be
considered resolved.

\subsection{Defining the Sample}

In this section, we describe how we defined a sample of candidate
clusters based on their individual properties and then how we
statistically corrected that sample in order to determine the number
of clusters in each galaxy.

We defined the sample of candidate clusters conservatively.  Using the
completeness data above, we interpolated the 90\% completeness
function for each galaxy to all core radii and used only clusters
brighter than this limit.  The limit corresponded to a V$_{\rm profile}$
between 24.15 and 24.55 (M$_{\rm V}$ of -7.0 to -6.6),  and was chosen to
ensure the accuracy of the measured properties of the cluster
candidates.  Clusters brighter than V$_{\rm profile}$=19.65
(M$_{\rm V}$=-11.5) were also rejected.  This limit on bright objects was
chosen to include even the brightest globular clusters.  It should
also include all but the most massive young clusters (e.g. all the
young and old clusters in the LMC would be included).  The final
number of cluster candidates in each galaxy is shown in Table~2.

It is important to consider the possible sources of contamination in
our sample.  
\begin{enumerate}
\item Background galaxies - especially compact ellipticals and bulge components of spiral galaxies.
\item Foreground stars - faint stars in our own Galaxy.
\item Compact, symmetrical HII regions in the target galaxies.
\item Single, very bright stars in the target galaxies.
\item Spurious detections in high surface brightness regions with
  many overlapping sources.
\end{enumerate}

Visually examining the data, only one of our galaxies (VCC 1374) contained
problematically high surface brightness regions.  The sources detected
inside two high surface brightness
regions in that galaxy were removed from the sample.  These two
regions can be
clearly seen in Figure~14 outlined by white boxes.  
The contamination of the sample by
extremely bright stars was minimized by
our inclusion of only those objects brighter than M$_{\rm V}\sim-7$.
Contamination of our sample by compact HII regions will be examined
later using the H$\alpha$ data in a subsample of the galaxies.

We obtained background and foreground (BG/FG) source counts from the
flanking (WF2 and WF4) fields.  In this section we describe how we
found the expected BG/FG level for each galaxy.  The
expected BG/FG levels are then used in two ways: (1) to remove the
BG/FG sources from the sample of candidate clusters as a whole,
described in \S4, and (2) to determine the number of red candidate
clusters in each galaxy for use in computing individual specific
frequencies as described in \S4.1.  

We based the expected BG/FG level on the mean level of sources in the flanking
fields in the entire sample.  However, because each galaxy has
different completeness limits, the BG/FG level varies from galaxy to
galaxy.  To determine the BG/FG level in a specific galaxy, all the
sources in the flanking fields were put through the 90\% completeness
cut and bright-end V$_{\rm profile}>$19.65 cut as described earlier
in this section. Then the BG/FG level and error were determined by
taking the mean and standard deviation of the number of remaining
sources in each flanking field.  The values for each galaxy are shown
in Table~2.  Note that the number of cluster candidates in Table~2 is not
corrected for the BG/FG level in any way.  We also note that because
the Virgo and Fornax fields  had nearly identical average BG/FG counts
and standard deviations, all fields were used in the BG/FG level
determination.   

Figure~6 shows the number of candidate clusters vs. the number of
BG/FG sources on each chip.  About half of all the galaxies, and 9 of
the 11 galaxies with confirmed cluster membership, appear to have
significant number of candidate clusters.  The remaining galaxies have
a number of candidate clusters consistent with the BG/FG contamination
level.  

We note that in the tables, figures, and analysis that follow we
excluded the eight 
candidate clusters with R$_{\rm core}$ values of 20pc.  These objects were
either the bulges of obvious background galaxies, or were in highly
confused regions and most likely have inaccurate sizes, magnitudes and
colors.   

\subsection{H$\alpha$ Observations}

For the seven galaxies that are confirmed members of the Virgo cluster
we have deep H$\alpha$ observations.  The reduction of these data was
described in \S2.  The H$\alpha$ data 
were used to find the galaxy-wide H$\alpha$ flux assuming a distance
of 17 Mpc (Table~3).  From the H$\alpha$ fluxes, we
derived star formation rates using the relationship given by
\citet{kennicutt94}:
\begin{equation}
SFR [\frac{M_{\odot}}{yr}] = \frac{L_{H\alpha}}{1.25 \times 10^{41}
  [ergs/sec]} 10^{0.4 A_{H\alpha}}
\end{equation}
where $L_{\rm H\alpha}$ is the H$\alpha$ luminosity and $A_{\rm H\alpha}$ is
the absorption by dust at H$\alpha$ (0.78 A$_{\rm V}$).  The values
for the star formation rates given in Table~3 
were calculated assuming no extinction by dust internal to the
galaxies and are thus lower limits.  Assuming absorption similar to
that of hot stars in the LMC, $A_{\rm H\alpha} \sim 0.3$
\citep{zaritsky99}, neglecting extinction 
would make the SFRs in Table~3 
roughly 30\% too small.  Note that the H$\alpha$
fluxes and SFRs in Table~3 have errors dominated by the absolute
calibration of the photometry (i.e. 50\%).  The H$\alpha$ fluxes are
in rough agreement with those given in \citet{heller99} for VCC 
328, 1374 and 1992.  The largest difference is 
for VCC 1374, where our flux is roughly a factor of two (2$\sigma$)
smaller than in \citet{heller99}.  We  
also detect H$\alpha$ in VCC 1822 where they had a just an upper limit
of $\sim 10^{38}$ ergs/sec.  The values found for star
formation rates fall in the same range as the dwarf irregulars in the
Local Group \citep{mateo98}.  

As mentioned in \S3.4 compact HII regions are one possible source of
contamination in our sample.  To
determine the possible contamination of 
clusters by HII regions, we first measured the H$\alpha$ flux in a
1$\arcsec$ annulus around the location of each cluster.  Then, using a
Balmer decrement of 2.8 \citep{osterbrock89} and the WFPC2/F555W filter
transmission curve ($\sim$7.5\% at H$\beta$ and $\sim$1.3\% at H$\alpha$) we
calculated 
the contribution of the H$\alpha$ and H$\beta$ line
flux to the V-band magnitude, V$_{\rm H\alpha}$.  The relation between
this derived magnitude and 
the actual contribution to the F555W flux is somewhat uncertain due to
three effects: (1) the absolute calibration of the H$\alpha$ is
reliable only to about 50\%, (2) any dust will reduce the H$\beta$
flux relative to the H$\alpha$ flux and (3) the poorer resolution of
the ground based H$\alpha$ data means that any H$\alpha$ flux located
within 1$\arcsec$ of the cluster candidate will be included in our
value for V$_{\rm H\alpha}$.
Taking into account our errors, we considered a cluster
candidate a possible HII region contaminant if it had a V$_{\rm H\alpha}$
brighter than, or  
within 0.5 magnitudes of, its V$_{\rm profile}$.  Our detection limit of
$\sim 3 \times 10^{35}$ ergs/sec results in V$_{\rm H\alpha}$ $\sim$ 26, and is
therefore sensitive enough to detect contamination in even our faintest
sources.  Only 32 out of 185 cluster candidates
were found to be contaminated by H$\alpha$ emission, 
with 27 of those
lying in the two galaxies
brightest in H$\alpha$, VCC 1992 and VCC 1374.   Figure~7 shows the
properties of the H$\alpha$-contaminated candidate clusters.
Of the blue cluster candidates (V-I $<$ 0.85), $\sim$26\% were found
to be likely HII regions, while only 5 (6\%) of the red candidate
clusters were contaminated by H$\alpha$, all located in VCC 1374.  
This difference between blue and red cluster candidates is not
surprising, since HII regions are expected to appear blue through the
F555W and F814W filters.  This suggests that our sample of red
clusters, which we consider to be candidate globular clusters, are
essentially uncontaminated by HII regions.

\section{Properties of the Cluster Sample}

Of our galaxy sample, only 11 galaxies, 7 in Virgo and 4 in Fornax,
have known velocities that are less than 3000 \kms (see Table~1).
These are the only galaxies for which cluster membership is confirmed and
therefore to which the distance is known.  We note that most of the
other galaxies in our sample (including three VCC galaxies with high
velocities) have insignficant or barely significant  
numbers of candidate clusters, as might be expected for more distant
galaxies with fainter globular cluster systems, or for fainter galaxies
within the Virgo and Fornax groups.  In this
section, we consider the bulk properties of the cluster candidates
using {\it only the 11 likely cluster members}, FCC 76, 120, 247 and
282 and VCC 83, 888, 1374, 1448, 1822 and 1992.  These 11 galaxies
contain 237 cluster candidates and an expected BG/FG contamination
(see Table~2) of 51 sources.  

Figure~8 shows the properties of the candidate clusters before (thin) and
after subtraction of the BG/FG (thick).  This subtraction was done by
selecting the number of expected background sources at random from the
BG/FG population and then removing the closest matching cluster
candidate (from the sample as a whole) in
V$_{\rm profile}$, V-I and R$_{\rm core}$.  Error bars on the histograms of
the subtracted sources were determined by taking the standard
deviation of 200 Monte Carlo simulations in which the candidate
cluster list and the BG/FG list were drawn at random from the initial
sample.  The most notable difference between the BG/FG sources and the
candidate clusters is in color - the BG/FG are mostly red
objects, and account for almost all of the candidate clusters with
colors of V-I $>$ 1.6. The BG/FG also contains many compact (R$_{\rm core} <
1$ pc) objects, as would be expected from foreground stars.  The
distribution of cluster candidates peaks at V-I $\sim$ 1 and R$_{\rm core}
\sim$ 0.  The leftmost bin in V$_{\rm profile}$ ($>$ 24 mag) is strongly
affected by our completeness-based cuts and therefore the apparent
turnover at faint magnitudes is not significant.  However, a turnover
at that magnitude would be expected for the typical globular cluster
system with M$_{\rm V,peak}\simeq$-7.4.  

We can divide the sample based on color to attempt to separate old
from young clusters.  The single stellar population (SSP) models of
\citet{girardi00} show that, for any metallicity, clusters with V-I of
$>$ 0.85 are likely to be older than 10$^9$ years, while clusters bluer
than this are likely to be younger.  Using this cut, we regard the
redder/older clusters as globular cluster candidates.  Even if we are
detecting clusters with ages of a few billion years, with our
magnitude cutoff (M$_{\rm V} \lesssim -7$) these clusters will have
globular cluster masses.  The bluer objects are most likely
young open or massive clusters, with a small fraction being HII
regions (see above).  Table~3 gives the numbers of blue and red
cluster candidates in each galaxy, while Figure~9 shows histograms comparing
the magnitudes and sizes of all 114 blue (thick) and 72 red (thin) 
cluster candidates left after BG/FG subtraction.  The blue objects
tend to be dimmer and more compact
than the red objects.  This result can be explained by either the presence
of single bright stars or compact open clusters among the blue cluster
candidates.  

Figure~10 shows the radial distribution of the blue
vs. red cluster candidates after BG/FG subtraction.  The red cluster
candidates typically lie 
farther from the center of the galaxy than the blue cluster
candidates.  K-S tests show that the distributions are significantly
different, with P values ranging between 10$^{-5}$ and 10$^{-3}$
depending on the background subtraction.  This distribution is what
would be expected if the red cluster candidates are an extended
population of globular clusters while the blue cluster candidates are
younger clusters formed in the body of the galaxy.  

The color and magnitude information also allows us to roughly
approximate the masses of our candidate clusters.
Figure~11 shows a 
color-magnitude diagram of background-subtracted cluster candidates
with lines representing the color-magnitude evolution of single
stellar populations of three masses \citep{girardi00} ranging in ages
between 60 Myr and 18 Gyr.  The blue
clusters almost entirely lie inbetween masses of 10$^4$ and 10$^5$
M$_\odot$, while the redder clusters all have masses significantly
greater than 10$^5$ M$_\odot$, and thus do represent objects with
masses similar to those of globular clusters.

The Milky Way globular cluster system (MWGCS) \citep[][revised 2003]{harris96}
provides a useful point of 
reference with which to compare our globular cluster candidates.  In
order to facilitate this comparison, we took the \citet{harris96}
catalog data and passed it through our completeness-based cuts (see
\S3.4) assuming a distance modulus of 31.15.  Of the 141 clusters 
with M$_{\rm V}$ and R$_{\rm core}$ measurements in the catalog, 83
survived the cuts.  Of the 96 
that also had V-I color information, 76 survived the cuts.  We then
compared the
surviving Milky Way clusters to our BG/FG subtracted sample of
candidate clusters.  Figure~12 shows histograms comparing the
V magnitudes, V-I colors and R$_{\rm core}$ of the MWGCS to the 72
cluster candidates with V-I $>$ 0.85.  
These plots exhibit some notable similarities and differences between
our red cluster candidates and the MWGCS.  

First, the 
luminosity function of the red
cluster candidates is similar to that of the MWGCS, with a K-S P=0.68.
This indicates that {\it the luminosity function of the red candidate
clusters is consistent with the 'Universal' globular cluster
luminosity function} found in other galaxies \citep{ashman98}.  This
similarity also supports the idea that our red candidate clusters are,
in fact, globular clusters.

Second, we consider the color distribution.  The MWGCS V-I
distribution is shown both before and after subtraction of the
foreground reddening, where E(V-I)$= 1.35$ E(B-V) \citep{cardelli89}.
In both cases, there is a clear color cut off at V-I $\sim$ 0.8,
similar to the color cut we used to select our candidate
clusters.  All the galaxies in our sample have foreground
reddenings E(B-V) $<$ 0.04 \citep{schlegel98}, with an average of
E(B-V) = 0.022.  However, the internal reddening due to dust is likely to be
larger for clusters seen inside / behind the body of the galaxy.
Corrected only for the mean foreground reddening, the median color of
our  sample is  V-I $= 1.00$, corresponding to an [Fe/H] of -1.2 using
the relation from \citet{kundu98}.  This median color is redder than
the 0.91 of the dereddened MWGCS, but is less than the average V-I of
1.04 for elliptical galaxies \citep{kundu01}.  The median color is
also redder than the globular cluster systems in Virgo, Fornax and Leo
cluster dwarf elliptical
galaxies \citep{lotz03}, which are found to have a correlation between
host galaxy luminosity and redder colors / higher metallicity.
The unknown internal reddening prevents us from determining if their
is a similar correlation in our galaxy sample.  Because of the
unknown reddening, our median color is a redward limit and the metallicity 
estimate is an upper limit.  This suggests that the clusters in
the dwarf galaxies in our sample are in general metal-poor.  

Finally,
 the bottom panel of Figure~12 shows that the distribution of
core radii of the red  
cluster candidates appears to be broader than that of the MWGCS.
However, errors in the determination of the core-radius could make the
distribution appear broader than it is.  To compensate for this, we
took the detected Milky Way clusters and added to each a random error,
the magnitude of which was determined by its core radius.  This error
was interpolated from the median of the errors shown in Figure~4b.  A
K-S test was performed using the resulting distribution of Milky Way
core radii and that of the Virgo/Fornax red cluster candidates.  
The random errors applied to the Milky Way data resulted in a range of P
values with a median value of 0.002 and a standard deviation of
0.009.  This test confirms that the red cluster candidates are
significantly broader than the Milky Way globular clusters.  Note that
the varying completeness limit inhibits our ability to detect broader
clusters relative to more compact ones, adding to the significance of
the difference between the cluster candidates and the MWGCS.

To summarize, comparison of cluster candidates with V-I $>$ 0.85 to
Milky Way globular clusters suggests that their luminosity functions are
similar, while the red cluster candidates have larger sizes.  The
average color is uncertain due to unknown reddening, but is consistent
with the red cluster candidates
being relatively metal poor.

\subsection{Specific Frequencies}

As discussed above, the number of red (V-I $>$ 0.85) cluster 
candidates, when corrected for the expected BG/FG level, provide
an estimate of the number of globular clusters in each galaxy in our
sample.   
A common way of expressing the number of clusters in a galaxy relative
to its luminosity is the quantity S$_{\rm N}$, the specific frequency
\citep{harris81}.
The specific frequency is defined as:
\begin{equation}
{\rm S_{N}= N_{GC} 10^{0.4({M_{V}+15})}}
\end{equation}
where N$_{\rm GC}$ is the number of globular clusters in a galaxy that has
absolute magnitude M$_{\rm V}$.  

In order to determine an accurate value for the specific frequency we
need to: (1) remove the BG/FG contamination from our number of red
candidate clusters, (2) correct for the unobserved portion of the
luminosity function, and (3) correct for the incompleteness in our
sample.  We remove the expected BG/FG level from our red cluster
candidate count in
a fairly simple way.  Of all the
BG/FG sources, 75\% of them have V-I $>$ 0.85.  Therefore we subtracted the
expected number of red BG/FG sources (75\% of the number given in
Table~2) from the number of red candidate clusters (Table~3) to obtain
N$_{\rm GC}$.   
Using this value for N$_{\rm GC}$ we derived a
conservative estimate of the specific frequency, shown as S$_{\rm N,min}$ in
Table 3.  We consider this estimate conservative because it is
uncorrected for both steps (2) and (3) above. 

Traditionally, S$_{\rm N}$ values are corrected for any unobserved
portion of the globular cluster luminosity function.  Given a Milky 
Way globular cluster luminosity function, 
with M$_{\rm V,peak}$=-7.33 and $\sigma$=1.23 \citep{ashman98}, we
determined the fraction of sources fainter than the 90\% completeness
limit for point sources in each galaxy, and then scaled up N$_{\rm GC}$ by
this factor.  The resulting luminosity function corrections have 
values of $\sim$1.4.  These corrections and the specific frequencies
calculated with them (``S$_{\rm N}$ w/ Corr'') are
shown in Table~3 and plotted in Figure~13.  

We are unable to correct
for the incompleteness in our data because
the incompleteness is dependent on the unknown size distribution of the
clusters.  However, given that most of
our sources are fairly compact, and that we define our sample using
only clusters brighter than
the 90\% completeness level, the correction for incompleteness should
be small.  

In both S$_{\rm N}$ calculations, M$_{\rm V}$ was derived from the
magnitude in Table~2 assuming a distance modulus of 31.15; the error
in the bars include the effects of Poisson noise, the errors in the
BG/FG counts and galaxy magnitudes ($\Delta$V$_T$ in Table~2), and
uncertainty introduced by the errors in the V-I colors.   

Our calculated specific frequencies may be in error due to the following
effects: (1) Internal reddening in the
galaxies can cause some young clusters to have colors that would land
them in our red sample.  This would result in an overestimate of the
specific frequency.  Given the low star formation rates of most of our
galaxies, it is unlikely that there are large numbers of young
embedded clusters in our sample.  Such clusters might also
be expected to be associated with H$\alpha$ emission.  Of the seven
galaxies in our H$\alpha$ sample, only one (VCC 1374) had any red
clusters associated with H$\alpha$ emission.  The effect of the random
errors in V-I color on the specific frequency are included in the error
estimates given in Table~3 and Figure~13.
(2) The lack of a completeness correction.  In both estimates of the
specific 
frequency given above, we are unable to correct for the possible
presence of faint, extended objects.  This could result in an
underestimate of the specific frequency.
(3) In order to compare
our S$_{\rm N}$ values with those of early-type galaxies it is traditional to
``age-fade'' the magnitude of the galaxies to the magnitude it would
have if it were an old stellar population \citep[][p. 310]{carney01}.
Using a scenario in which a galaxy forms stars at a constant rate for 5
Gyr and then fades for 5 Gyr, \citet{miller98} find a fading of
$\sim$1.5 magnitudes in V, resulting in factor of $\sim$4 increase in
specific frequencies.   
However, without better knowledge of the stellar populations of the
galaxies, and the destruction timescales for the globular clusters, it
is not possible to determine what the exact increase in specific
frequency would be.
(4) Intermediate-age ($>1$ Gyr)
massive clusters may be included in this sample. 
To get a sense of the number of intermediate age clusters we might
be detecting, we used the LMC as a case study.  The LMC provides a nearby
laboratory in which cluster ages are well known.  Using the old and
intermediate-age LMC clusters given in \citep{geisler97} and our detection
limits, we would detect $\sim$10 clusters as red candidate clusters.
Only half of these are bona fide globular clusters, with the other half
being clusters with ages of a few gigayears.  
So it appears very likely that the presence of intermediate-age
clusters contributes to the specific frequency values we measure.  
However, we note that intermediate-age clusters may also be included in
most other extragalactic specific frequency measurements, which
typically don't include a method for estimating the ages and
metallicities of the clusters.  Spectroscopic and IR observations done
by \citet{puzia02} and \citet{larsen03} show that intermediate-age
clusters do exist in some early type galaxies.  
We discuss the specific frequency values further in \S5.

\subsection{Individual Galaxies}

In this section, we consider the distribution of candidate clusters in
individual galaxies from our sample.  In general, we find that most
galaxies have cluster candidates situated both inside and outside
their apparent optical boundaries.  The candidates located on the
galaxies tend to be blue, while those around the periphery are more often red.
This can easily be seen in Figure~10 which shows a large difference in
radial distribution between the red and blue cluster candidates.  This
distribution is consistent with these galaxies having recent star
formation concentrated near their centers and a more widely
distributed older globular cluster population.  

Previous classifications of a number of the galaxies is somewhat
uncertain.  Of the sample of 11 galaxies (Table~3), 8 are clearly
dwarf irregulars, while two are most likely dwarf
ellipticals/spheroidals (FCC 247 and VCC 1448) and one (FCC 282) is a
high surface brightness object with some suggestions of spiral
structure, but would most likely still be classified as an irregular
galaxy.  Figure~14 shows CMDs for the cluster candidates in each
galaxy and images with the location of red and blue candidates denoted
with circles and squares, respectively.  Larger images of the galaxies
are available from the authors by request.

\noindent {\it FCC 76 -- }  this fairly high surface brightness
irregular galaxy has five blue and two red cluster candidates on its
face.  However, 
around the outskirts of the galaxy, five of the seven
cluster candidates are red.  There are also a fairly large number
of clusters found at large radii (5 red, 3 blue).  

\noindent {\it FCC 282 -- } as noted above, this galaxy has an unusual
morphology and was classified as an Im/dEpec by \citet{schroeder96}.
Towards the outer portion of the galaxy, tightly wound 
spiral structure is visible.  However there is no bulge component and
the high surface brightness central area is mottled without any
obvious peak.  If FCC 282 is at the distance of the Fornax cluster, as
suggested by its velocity, it would have an M$_V$ of about -17, 
making it similar in luminosity to the LMC.  Based on its appearance,
this galaxy is almost certainly not an early type galaxy; we classify
it as an irregular.  

Most of the cluster candidates are located around the edge of the
inner regions of the galaxy (Figure~14) and are predominantly blue.
Around the outskirts of the galaxy there are 5 red and 2 blue
cluster candidates.   

\noindent {\it VCC 83 -- } this galaxy appears 
to be sandwiched between a pair of galaxies.  Given the other two
galaxies' small size ($<$ 15$\arcsec$), high surface brightness, and
spiral appearance, it seems that they are background 
galaxies in chance aligment.  We have attempted to measure the
magnitude of VCC 83 without including those galaxies.

\noindent {\it VCC 328 -- } this galaxy has a very prominent system of blue
clusters.  The 
cluster candidates appear to be located preferentially along the
edge of the apparent optical boundary of the galaxy (particularly towards the
lower right side in Figure~14), a property 
shared by the H$\alpha$ distribution (which can be seen in Fig.~1).
Of the 14 clusters located 
inside the apparent optical boundary of the galaxy, only one is red, while two
appear to be HII region contaminants.  

\noindent {\it VCC 1374 -- } this galaxy contains active star formation and has
an apparent 
diffuse H$\alpha$ component. The 64 candidate clusters are the most
detected in any galaxy in our sample.  This is despite the exclusion
of two of the most crowded
regions of the galaxy (seen outlined by white boxes in
Figure~14) in which cluster candidates could not be reliably
identified.  There is a mixture of blue and red candidates on the
galaxy face suggesting 
either the presence of both young and older populations or large
amounts of reddening.  Of the red candidates, 20\% of them are found
to be possible HII regions.  Given the differing resolutions of the
ground-based and HST images, this could either be due to chance proximity
of older clusters with HII regions, or evidence for highly
reddened young clusters.  If the red cluster candidates are in fact 
globular clusters, then the globular cluster system appears flattened, as
is found in the LMC \citep[][p. 106]{vandenbergh00}.  Of nine
candidate clusters found around the outskirts of the galaxy, all but
one is red, perhaps 
tracing a halo component to the globular cluster system.

\noindent {\it VCC 1448 -- } this galaxy, a.k.a. IC 3475, is the most
well-studied galaxy in 
our sample.  The galaxy is large ($\sim$1.7'), has a low surface
brightness, is very HI and HII poor, and has smooth isophotes
\citep{vigroux86,knezek99}, all properties which would generally
classify it as a dE galaxy.  However the presence of
a central bar and a large number of 'knots' on the surface of the
galaxy might suggest the galaxy is a dIrr.  All the 'knots' that
appear in our field of view appear to be candidate clusters.

Our observations cover the bar of the galaxy, but clearly do not cover the
entire halo.  Therefore, our integrated magnitude for the galaxy
predominantly represents the magnitude of the bar; it is more than a
magnitude fainter than the RC3.9 value of V$_T$=13.11
\citep{devaucouleurs95}.  This means that the specific frequency
listed in Table~3 is a measure of the 
local specific frequency, because it excludes much of the
halo light and any halo clusters that may exist.  Some bright cluster
candidates outside our field of view are clearly visible in Fig. 1e of
\citet{knezek99}.  

The 26 candidate clusters in VCC 1448 all fall between V-I of 0.77 and
1.30 (with the exception of one likely FG star at
V-I=2.35)  as can be seen in the color-magnitude diagram for VCC 1448
shown in Figure~14. This is consistent with the suggestion of \citet{vigroux86}
that the 'knots' were intermediate-age clusters, but is also
consistent with the clusters being a metal-poor globular cluster
system.  \citet{knezek99} find VCC 1448 to be 0.4 magnitudes bluer
than a typical dE and suggest that it is a metal-poor galaxy which has
recently finished its last episode of star formation.  Spectroscopic
observations or accurate three color photometry 
\citep[as in][]{puzia02} of these clusters 
would enable the determination of ages and shed some light on the star
formation history of this unusual galaxy.  

\noindent {\it VCC 1992 -- } is a diffuse irregular galaxy with a number of
prominent star formation regions with strong H$\alpha$ emission.  Eight
(more than half) of the blue cluster candidates are likely HII region
contaminants.  Three of the seven red cluster
candidates lie near the center of the diffuse emission, while three
others around the outskirts of the galaxy.  None are associated with the star
forming regions, suggesting that they are, in fact, old or
intermediate-age clusters.

\section{Discussion}

\noindent {\it Specific Frequencies -- } Based on our results, it
seems that dwarf irregular galaxies have 
a wide range of specific frequencies, including values greater than
five.  In the Local Group, the specific frequencies for many of the
irregulars are 0 (e.g. IC 1613, Sextans B) while the Magellanic
Clouds have S$_{\rm N}\sim$0.5 \citep{harris91} and WLM and NGC 6822 both
have one detected globular cluster, giving them S$_{\rm N}$ values of
$\sim$2 \citep{mateo98}.  So it appears from the Local Group that
dwarf irregulars typically have S$_{\rm N}$ values that range between
0 and 2.  This
is similar to the values obtained for spiral galaxies
\citep{carney01}.  

Although there are many uncertainties in our
specific frequency values (see \S4.1), it appears that the specific
frequencies found here for dwarf irregular galaxies VCC 83, 1374, 1992 and
FCC 76 and 120 (we omit likely dE galaxy VCC 1448 from
this list) are higher than their Local Group counterparts.  This result 
suggests that environment may play an important role in determining
the characteristics of a globular cluster system.  
Specifically, the high specific frequencies we observe might
be related to the high density environments that these Virgo and
Fornax dwarf irregulars live in.  This is a plausible effect given the
apparent connection between galaxy interaction and cluster formation
\citep[][p. 104]{ashman98}.   

The high specific frequencies of our sample also affect the
conclusions of \citet{miller98} on the possible
relationship of dE and dIrr galaxies.  \citet{miller98} found that the
nucleated dE galaxies 
have an average S$_{\rm N}$ of 7.5$\pm$1.8, while the average for
non-nucleated dE galaxies is 2.8$\pm$0.7.  They use this to suggest that the
dIrr galaxies (measured locally) 
did not have sufficiently high specific frequencies to evolve
into nucleated dE galaxies.  However, with age-fading, they suggested that 
dIrr galaxies could evolve 
into non-nucleated dE galaxies.  Our results modify this conclusion -
given the high S$_{\rm N}$ values we observe, it is
possible that both types of dE galaxies could evolve from dIrr
galaxies.

\noindent {\it Populous clusters -- } The best-studied irregular
galaxy, the LMC, 
is known to host a number of luminous, compact, blue clusters, often
called populous clusters that are not seen in our own Galaxy.  From
the \citet{bica96} catalogue, there are $\sim$ 35 objects with M$_V <$
-8.5 and colors bluer than B-V=0.5, which should be roughly
equivalent to our V-I=0.85 cutoff.  In our sample, 15 blue
candidate clusters are brighter than M$_V$=-8.5 (see Figure~11),
suggesting that a similar, but smaller, population exists among
the dwarf irregular galaxies in the Virgo \& Fornax Clusters.  Our sample
contains no blue objects as bright as R136.

\noindent {\it Size of red candidate clusters --} Figure~12 and our K-S
tests (see \S~4) indicate that the distribution of core radii of
our red cluster candidates is larger than the Milky Way globular
clusters.  There are also indications that globulars in the LMC, SMC
and Fornax are larger than in the Milky Way \citep{ashman98,vandenbergh91}.  
Using data from \citep{mackey03} for 12 globular clusters (with age
$>$ 10 Gyr) in the LMC, we find a mean and median core radius of 2.5
and 2.0 pc respectively.  For the Milky Way globulars in a comparable
range of absolute magnitudes (M$_V <$ -6.5), the mean and median core
radius is 1.5 and 0.9 pc \citep{harris96}.  
For elliptical galaxies, globular clusters have been found to
have a mean half light radius of 2.4 pc \citep{kundu01}, similar to,
and somewhat smaller than the value of $\sim$3 pc found for Milky Way 
globulars \citep{vandenbergh96}.  So it seems that irregular (and
perhaps dwarf ellipticals as well) have larger globular clusters than
the Milky Way and giant elliptical galaxies.  \citet{vandenbergh91}
suggests that the reason for the large sizes of clusters in the
Magellanic Clouds may be due to their formation in a lower density
environment.  However, destruction processes could 
also create differences in the average size of a population
\citep{fall77}.  It is 
therefore unclear whether the differences in size distribution are the
result of differences in formation environment or evolution of the
cluster system.  

\section{Conclusions}

We have studied the cluster systems of dwarf irregular galaxies using
HST/WFPC2 observations of 28 galaxies in the Virgo and Fornax
clusters, 11 of which are confirmed members.  In these 11 confirmed
members we have found 237 cluster candidates and determined their
magnitudes, core radii, and V-I colors.   Based on source counts of
neighboring, empty, WFPC2 fields, we expect a contamination of 51
background and foreground sources, which were statistically subtracted
from the data.  The cluster candidates were divided by color in order
to separate red (with age $\gtrsim$ 1 Gyr) and blue (age
$\lesssim$ 1 Gyr) cluster candidates.  We consider the red cluster
candidates to be globular cluster candidates.  The main results presented
here are:  
\begin{enumerate}
\item The red cluster candidates have the same magnitudes as expected
  for globular clusters.  More specifically, the luminosity function
  of Galactic globular clusters \citep{harris96}
  is well matched by the luminosity function of our red
  cluster candidates down to our 90\% completeness limit at M$_{\rm V}
  \sim$ -7.
\item The median color (V-I = 1.0) of the red cluster candidates is
  bluer than the median colors of globular cluster systems in
  elliptical galaxies \citep{kundu01}, even without a correction 
  for internal reddening.  This is consistent with the globular
  cluster systems in these dwarf irregular galaxies being metal-poor.
\item The distribution of core radii for the red candidate clusters is
  significantly larger than for 
  Galactic globular clusters.  Larger radii are also seen in the
  globular clusters of the LMC and other local group dwarves
  \citep{vandenbergh91} suggesting that this might be a general
  property of globular clusters in dwarf galaxies. 
\item The blue cluster candidates are in general more compact and
  fainter than the red cluster candidates.  Based on single-stellar
  population models, they have masses ranging between 10$^4$ and
  10$^5$ M$_\odot$.  H$\alpha$ data on 7 Virgo cluster galaxies
  suggests that 26\% of the blue cluster candidates are in fact HII
  regions. 
\item The specific frequencies for the galaxies exhibit a large
  scatter, but are surprisingly large, with 7 of the 11 galaxies
  having specific frequency values greater than two.  These values are
  surprising when compared to Local Group 
  dwarf irregulars and other nearby late-type
  galaxies, which have specific frequencies less than two.
  We note, however, that our specific frequency values are uncertain due to the
  possible presence of reddened young clusters, massive
  intermediate-age clusters and incompleteness.

\end{enumerate}

We plan to follow up this work with spectroscopic observations or
infrared imaging of some of these cluster systems.  These observations
of the cluster candidates will test the accuracy of our photometric
results by enabling accurate age and metallicity determination.  

This work was supported by the National Optical Astronomy Observatory,
which is operated by the Association of Universities for Research in
Astronomy, Inc., under cooperative agreement with the National Science
Foundation.  It also has been supported by the Gemini Observatory,
which is operated by the Association of Universities for Research in
Astronomy, Inc., on behalf of the international Gemini partnership of
Argentina, Australia, Brazil, Canada, Chile, the United Kingdom, and
the United States of America.  The authors would like to thank: Paul
Hodge, for his guidance and help observing, Peter Stetson for help
with DAOPHOT II, Eduardo Bica, for providing us with his tables, Nick
Suntzeff for his helpful comments, Ricardo Covarrubias, and Armin
Rest.

\clearpage

\begin{figure}[p]
\epsscale{1.0}
\plottwo{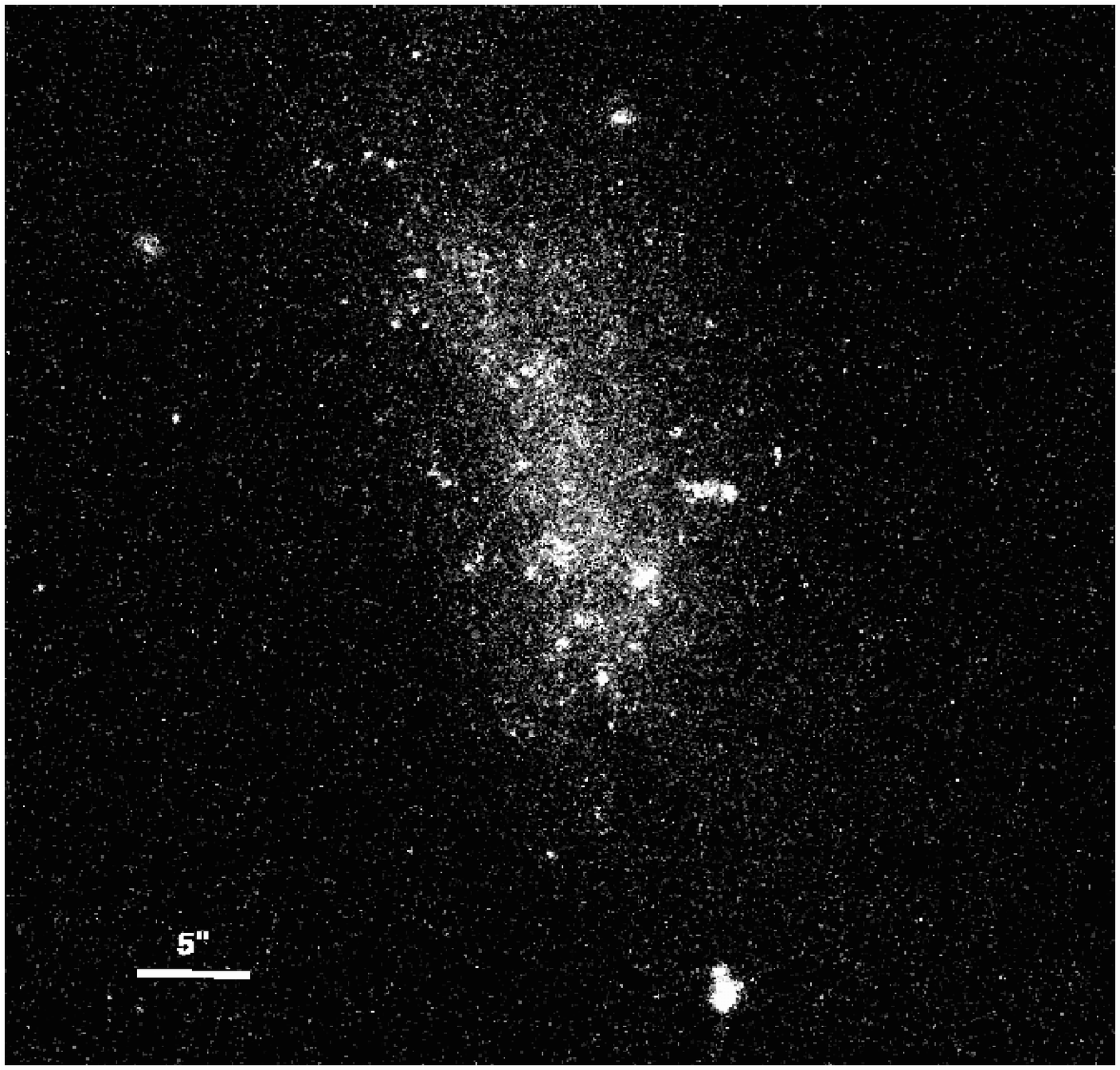}{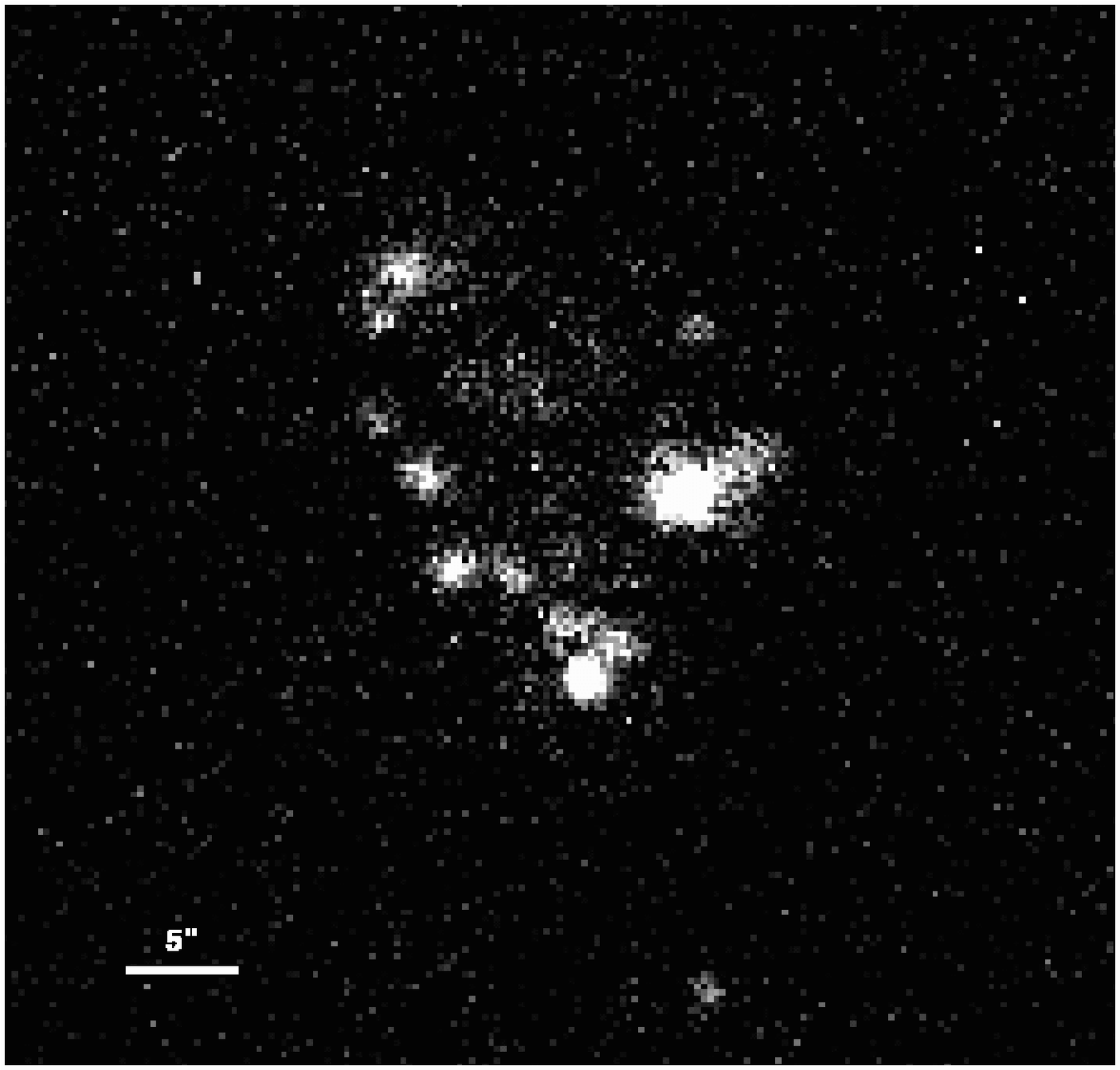}
\caption{on the left, (1a) shows the HST image of VCC 328, while on
  the right (1b) shows the H$\alpha$ image of the same galaxy.  The
  images are aligned and at the same scale.}
\end{figure}

\begin{figure}[p]
\epsscale{1.0}
\plotone{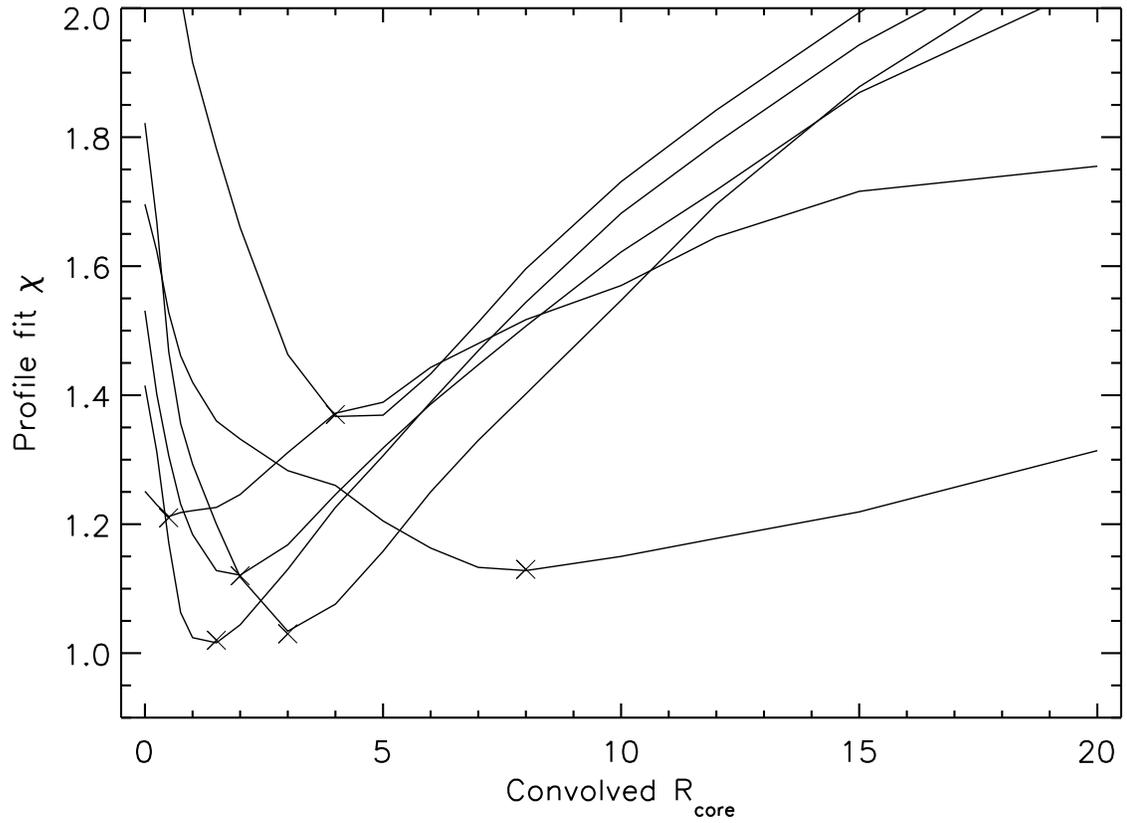}
\caption{The radius determination of six clusters with different
  R$_{\rm core}$ in VCC 1448.  The y-axis shows the $\chi$ values 
  (as output by DAOPHOT II)
  determined from fitting the model cluster profile to the object.
  The x-axis gives the core radius of the model cluster profile.}
\end{figure}

\begin{figure}[p]
\epsscale{1.0}
\plotone{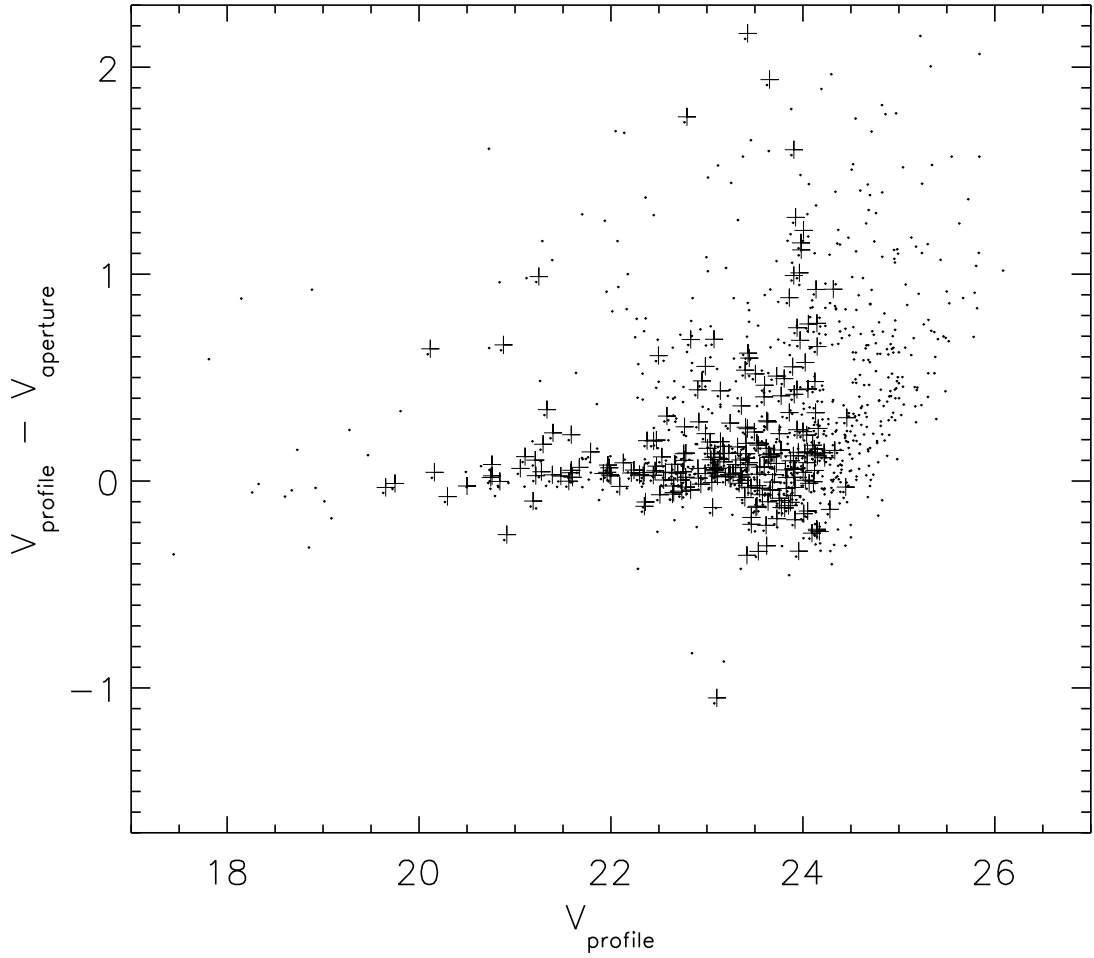}
\caption{A comparison of the photometry determined from the fitting of
model cluster profiles (V$_{\rm profile}$) to aperture photometry
(V$_{\rm aperture}$).  The dots represent all cluster candidates,
whereas the crosses denote the final sample of cluster candidates
in the confirmed Virgo and Fornax galaxies. }
\end{figure}

\begin{figure}[p]
\epsscale{1.0}
\plotone{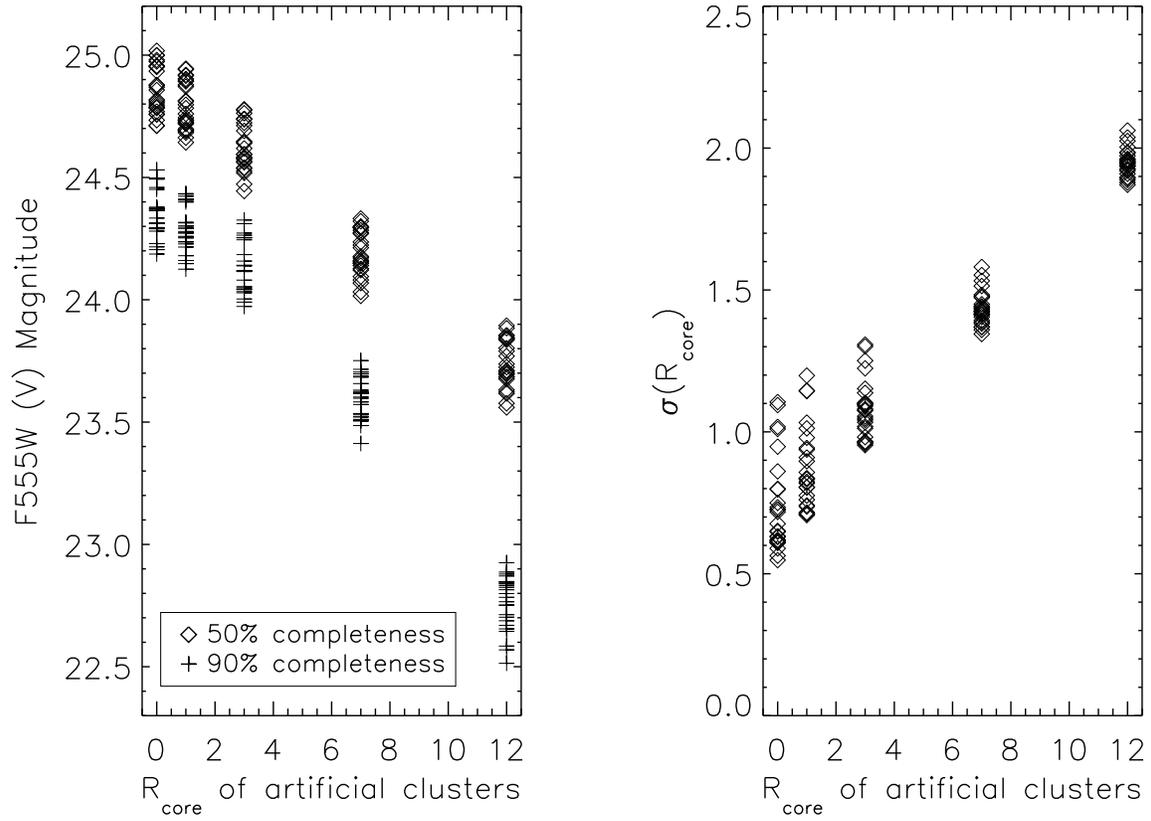}
\caption{{\it Left (4a)} The 50\% and 90\% completeness limits for all 26
  galaxies as determined from artificial cluster tests. {\it Right (4b)} The
  average error in core radius as a function of core radius for each
  galaxy.}  
\end{figure}

\begin{figure}[p]
\epsscale{1.0}
\plotone{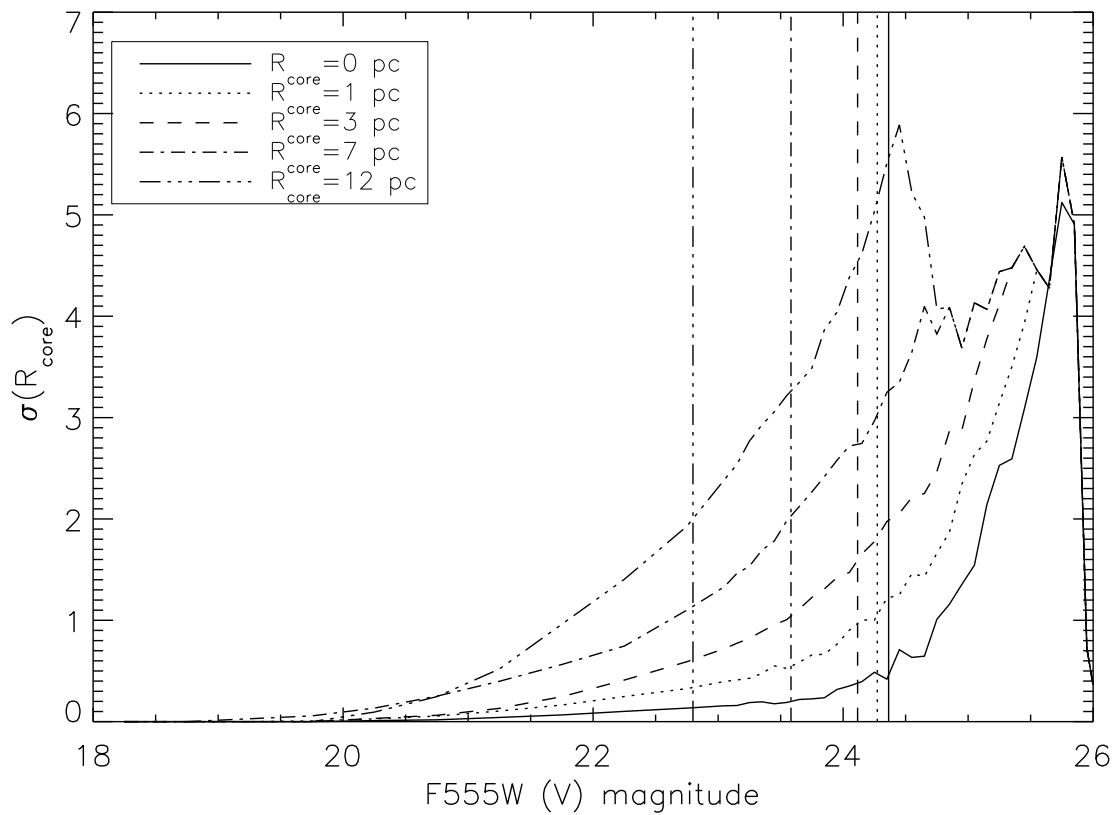}
\caption{Average errors in the core radius as a function of
  magnitude.  $\sigma$(R$_{\rm core}$) is the standard deviation of the
  absolute value of the measured core radius minus the expected
  core radius. The vertical lines show the 90\% completeness limits at
  each core radius.}
\end{figure}

\begin{figure}[p]
\epsscale{1.0}
\plotone{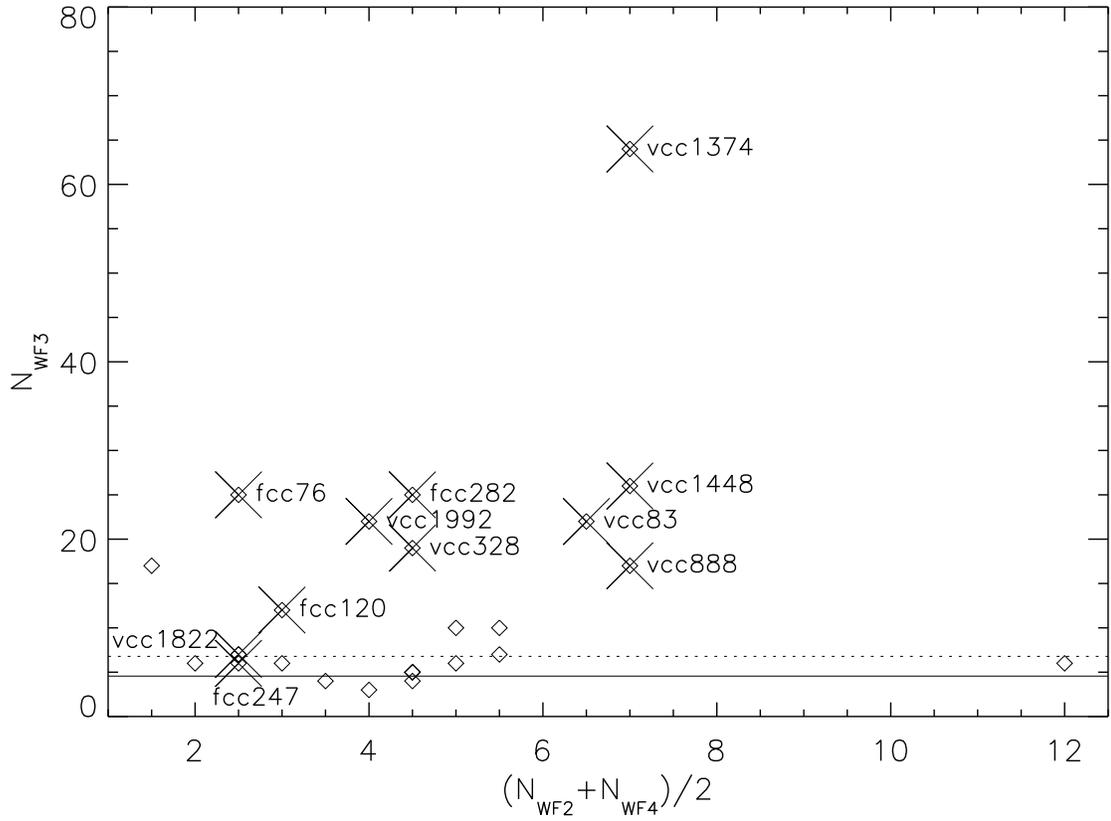}
\caption{The number of cluster candidates on the WF3
  chip versus the BG/FG levels on
  the other two chips for each galaxy.  Diamonds mark all the galaxies
  in our sample, while 
  the large X's mark those whose membership in Virgo and Fornax is
  confirmed.  The solid horizontal line shows the average BG/FG level,
  and the dotted line is 1$\sigma$ above the average level.}
\end{figure}

\begin{figure}[p]
\epsscale{1.0}
\plotone{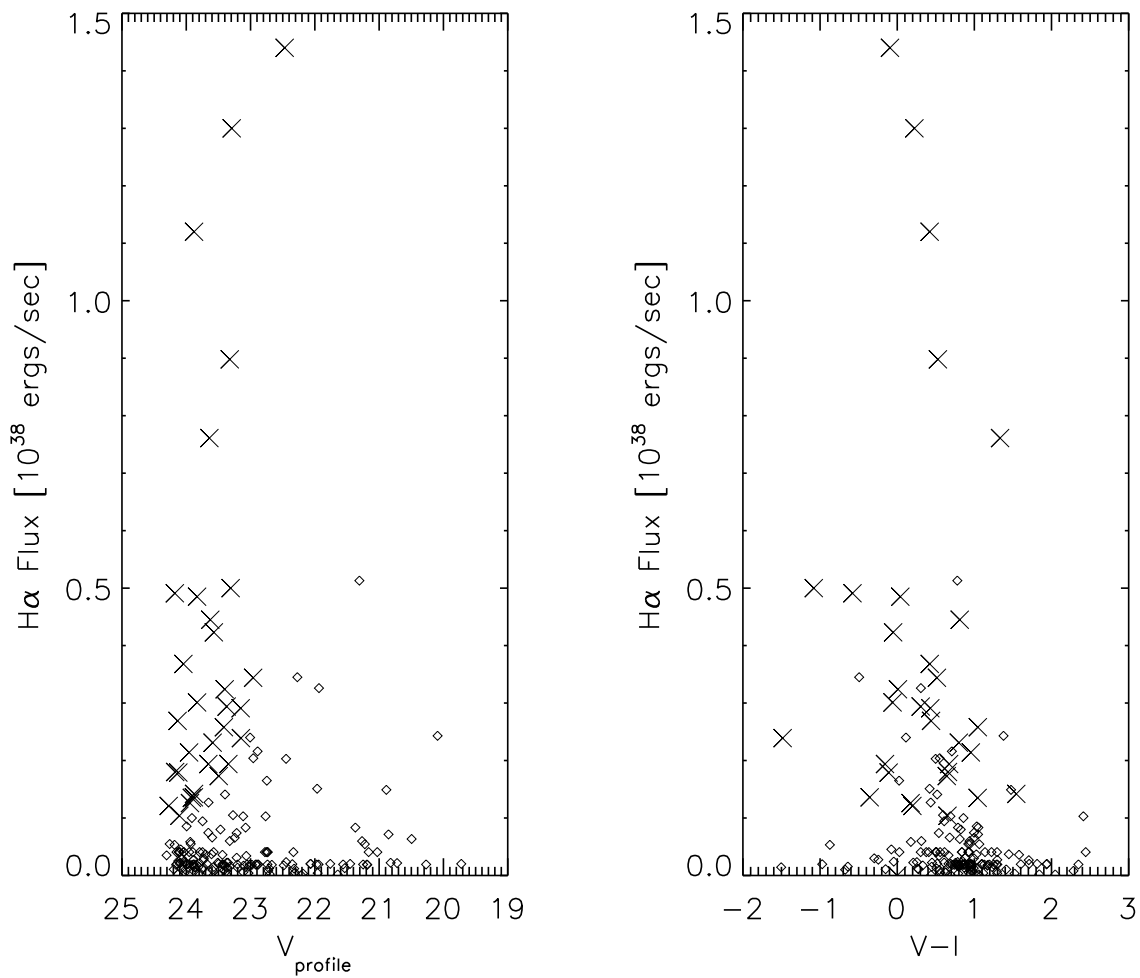}
\caption{Cluster candidates in the 7 Virgo galaxies are shown with
  their measured H$\alpha$ fluxes.  X's represent clusters candidates
  that we consider to be contaminating HII regions. On the left
  the V$_{\rm profile}$ magnitudes are shown, and on the right, V-I
  colors. }
\end{figure}

\begin{figure}[p]
\epsscale{0.7}
\plotone{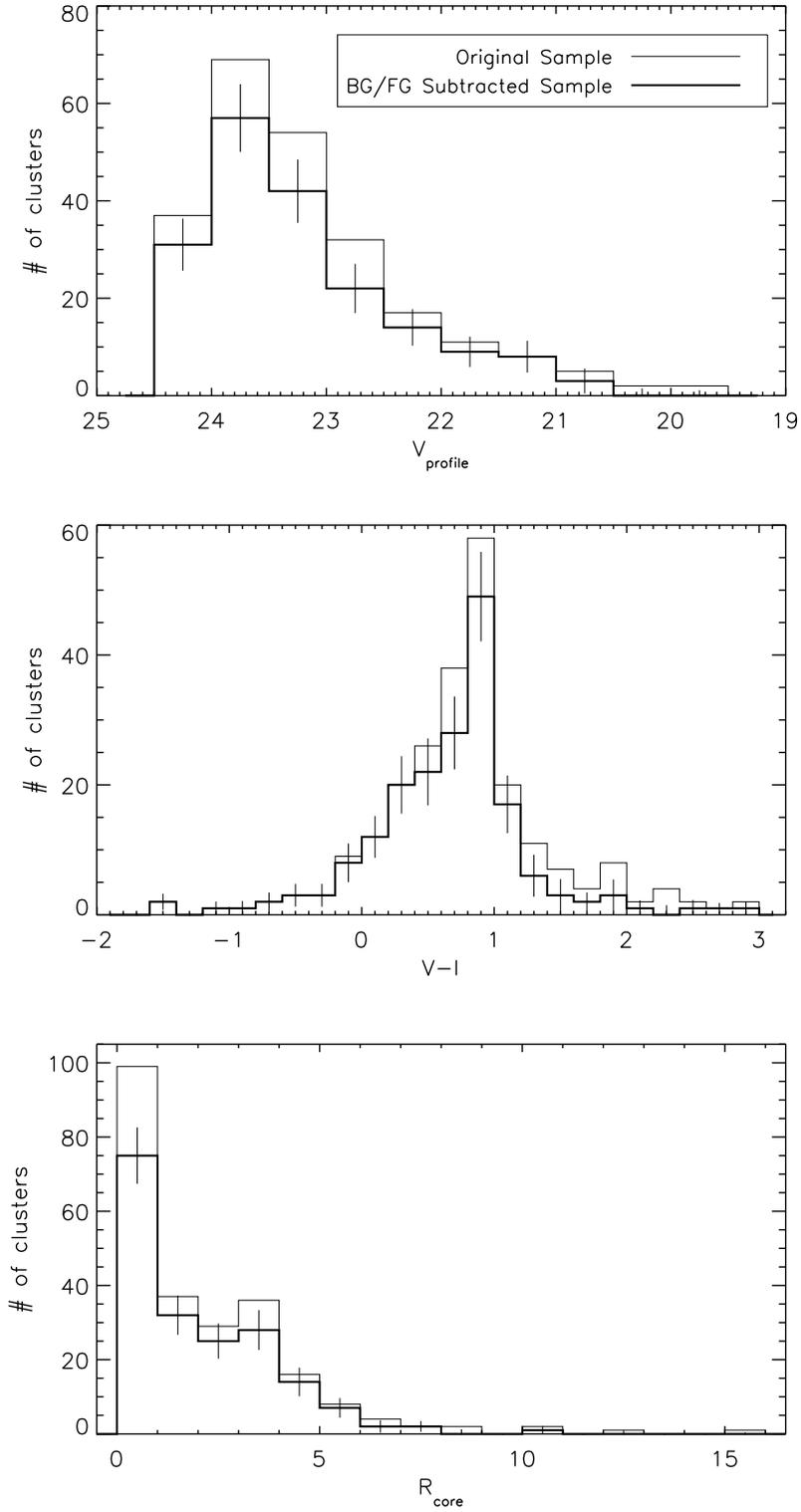}
\caption{Histograms showing the properties of the cluster candidates
  before and after BG/FG subtraction.  Error bars were determined
  using Monte Carlo simulations.  Only cluster candidates from the
  sample of 11 galaxies that are probable members of Virgo and Fornax
  are shown.} 
\end{figure}

\begin{figure}[p]
\epsscale{1.0}
\plotone{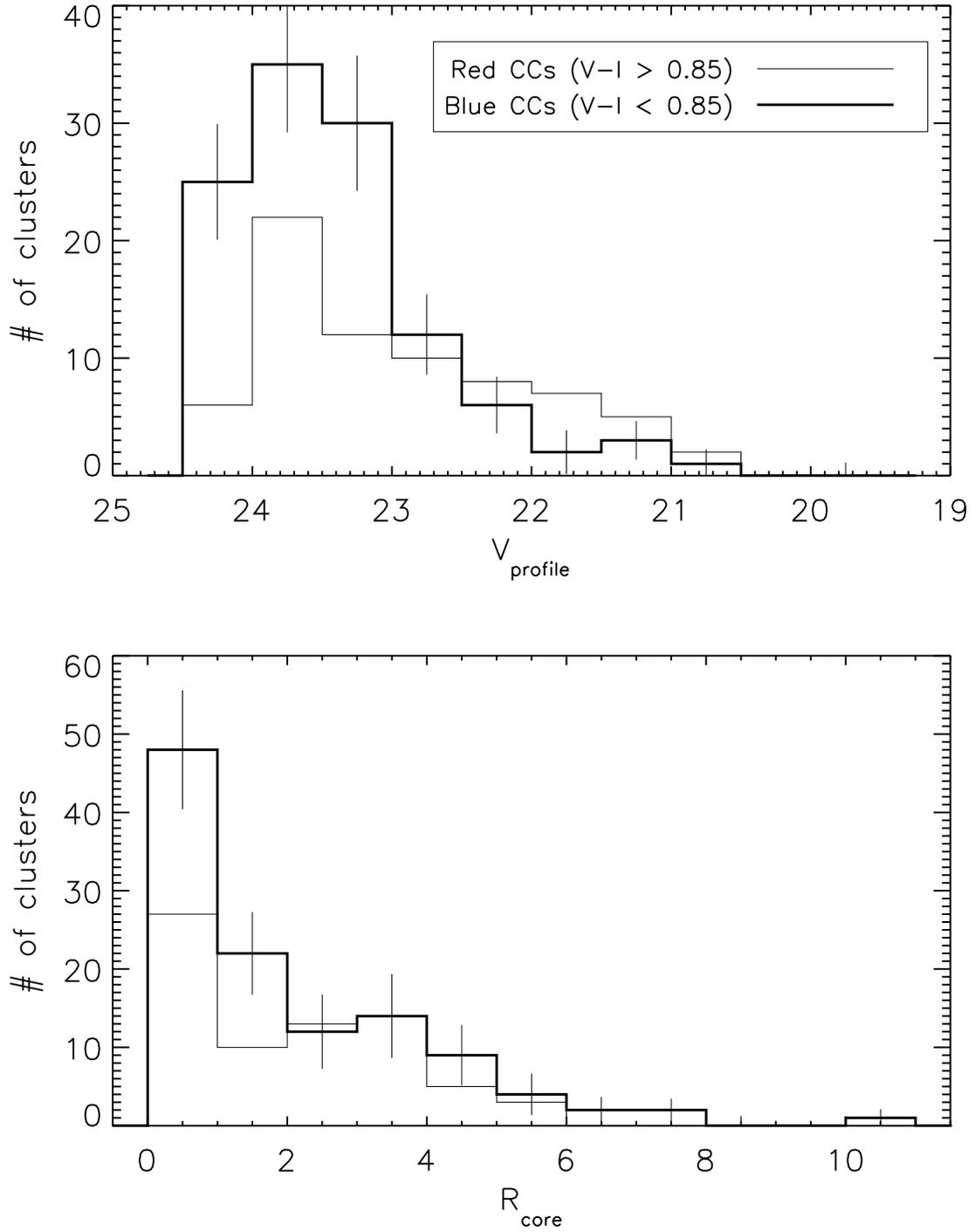}
\caption{A comparison of the magnitudes and core-radii of the BG/FG
  subtracted red and blue cluster candidates in the sample of 11 galaxies.}
\end{figure}

\begin{figure}[p]
\epsscale{1.0}
\plotone{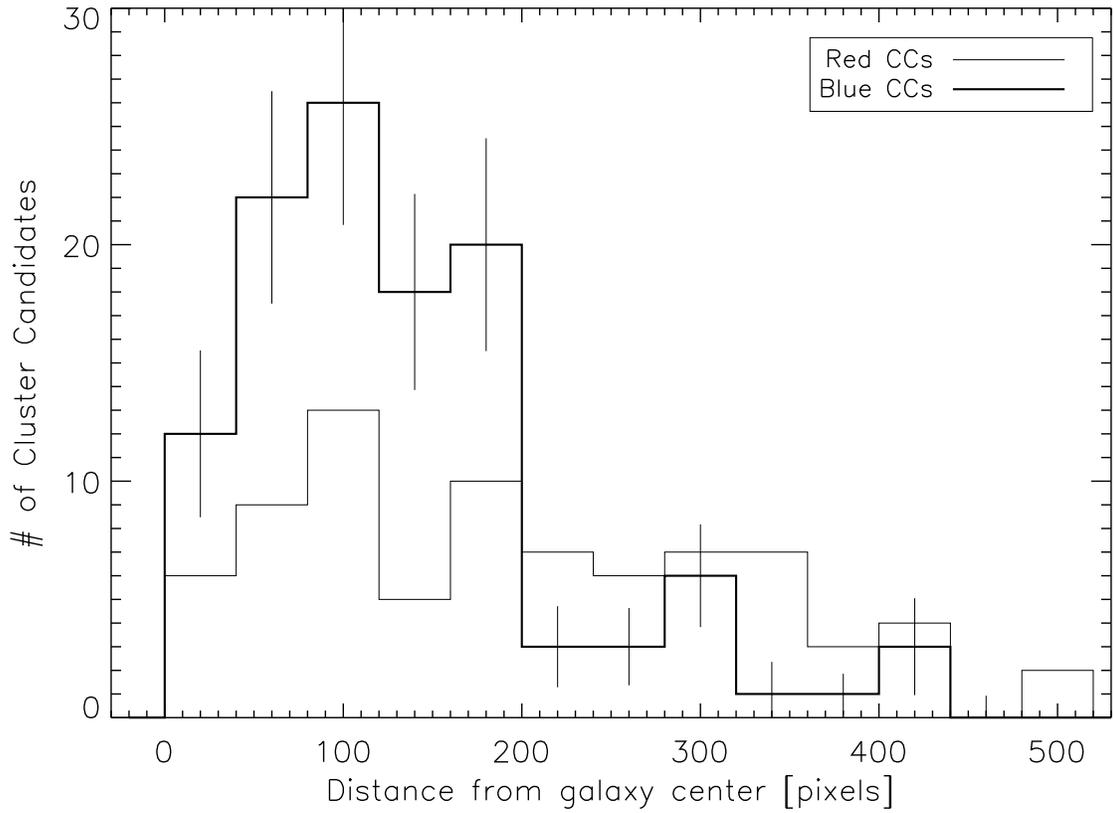}
\caption{The radial distribution of cluster candidates in the 11
  sample galaxies after BG/FG subtraction.  One pixel corresponds to
  0.1$\arcsec$, which at a distance of 17 Mpc corresponds to 8.2 pc.
  The blue and red samples are divided at V-I=0.85}
\end{figure}

\begin{figure}[p]
\epsscale{1.0}
\plotone{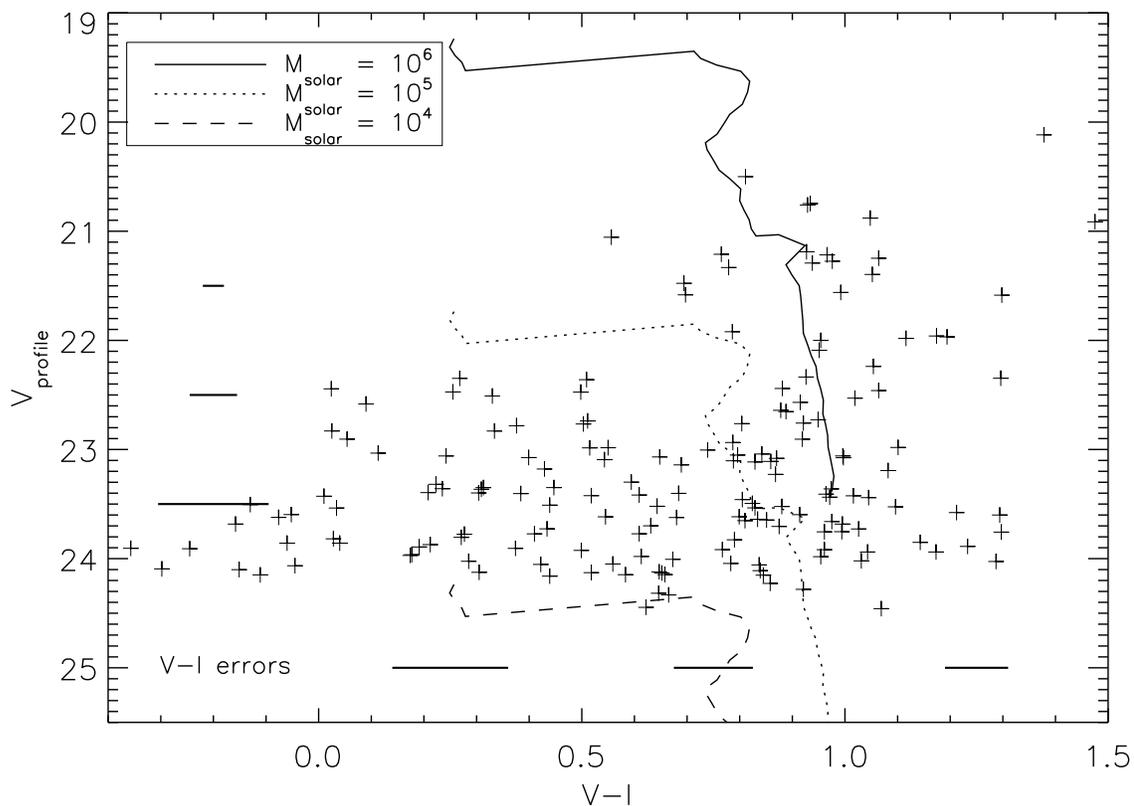}
\caption{A color-magnitude diagram of the cluster candidates from the
  11 sample galaxies.  Overplotted are [Fe/H]=-1.3 single stellar
  population models of 10$^4$, 10$^5$, 10$^6$ M$_\odot$ clusters from
  \citet{girardi00} with ages ranging from 60 Myr to 18 Gyr.  Bars on the bottom and left side display the
  median 1$\sigma$ 
  V-I errors between V-I of (0.0, 0.5), (0.5, 1.0) and (1.0, 1.5) and
  between V$_{\rm profile}$ of (21,22), (22, 23), and (23, 24)}  
\end{figure}

\begin{figure}[p]
\epsscale{0.7}
\plotone{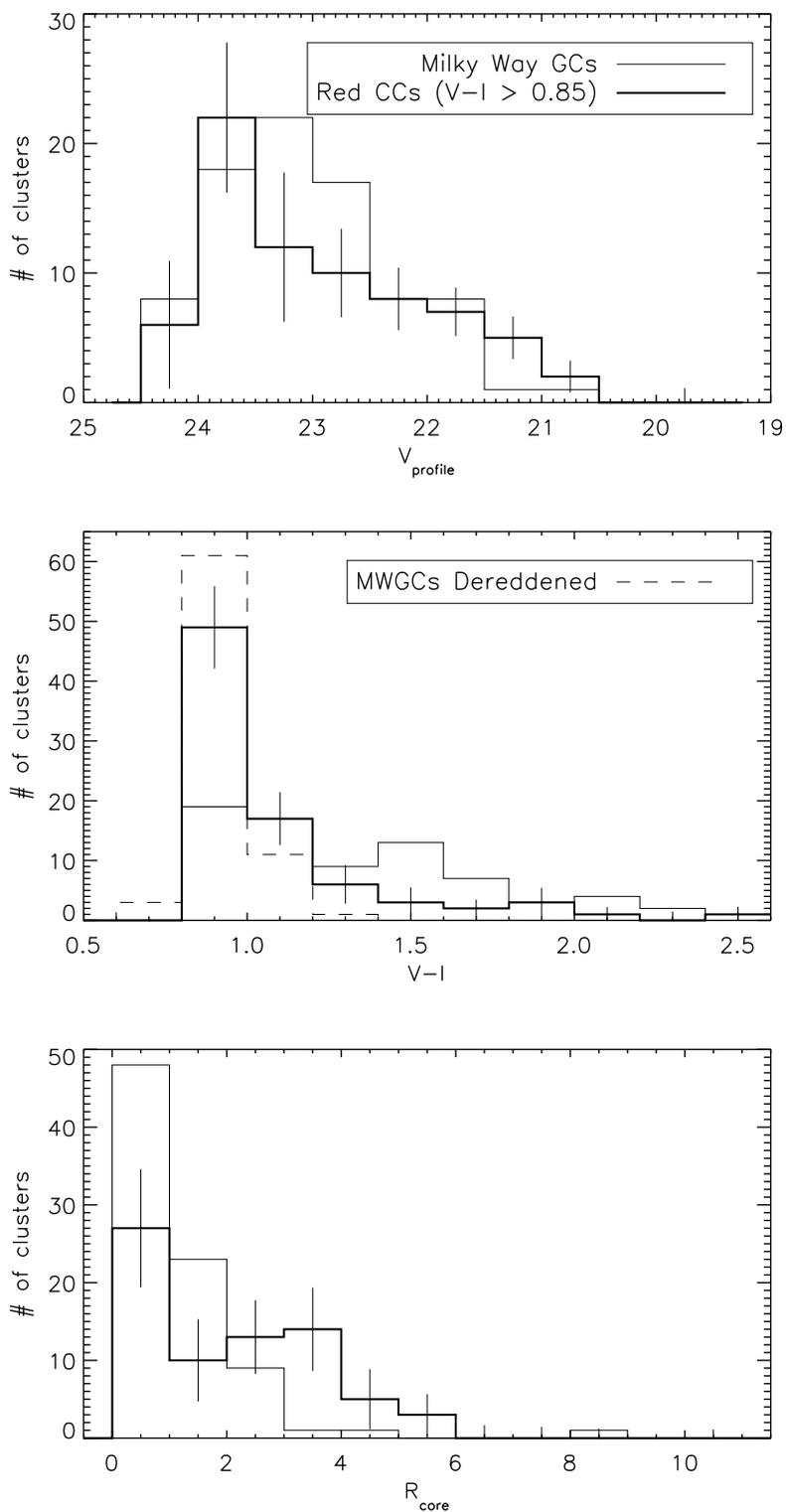}
\caption{A comparison of the magnitudes, colors and sizes of the
  red cluster candidates vs. the Milky Way Globular Clusters that
  would have been detected by our methods.  The cluster candidates
  shown in bold in each plot are only those with V-I $>$ 0.85.}
\end{figure}

\begin{figure}[p]
\epsscale{1.0}
\plotone{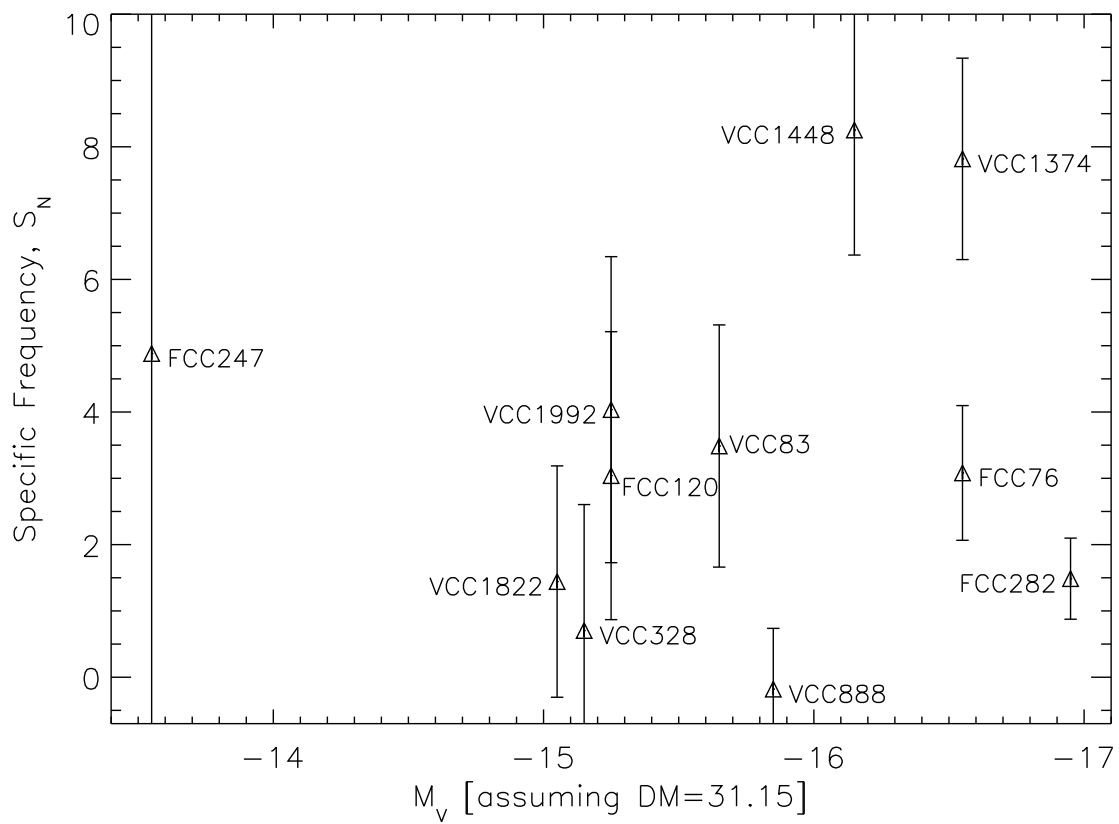}
\caption{The specific frequency for each galaxy as measured by the number of
  background subtracted, red (V-I $>$ 0.85) cluster candidates.  The
  specific frequencies plotted here are corrected for unobserved
  portions of the luminosity function as described in \S4.1.}
\end{figure}

\begin{figure}[tp]
\epsscale{0.66}
\plotone{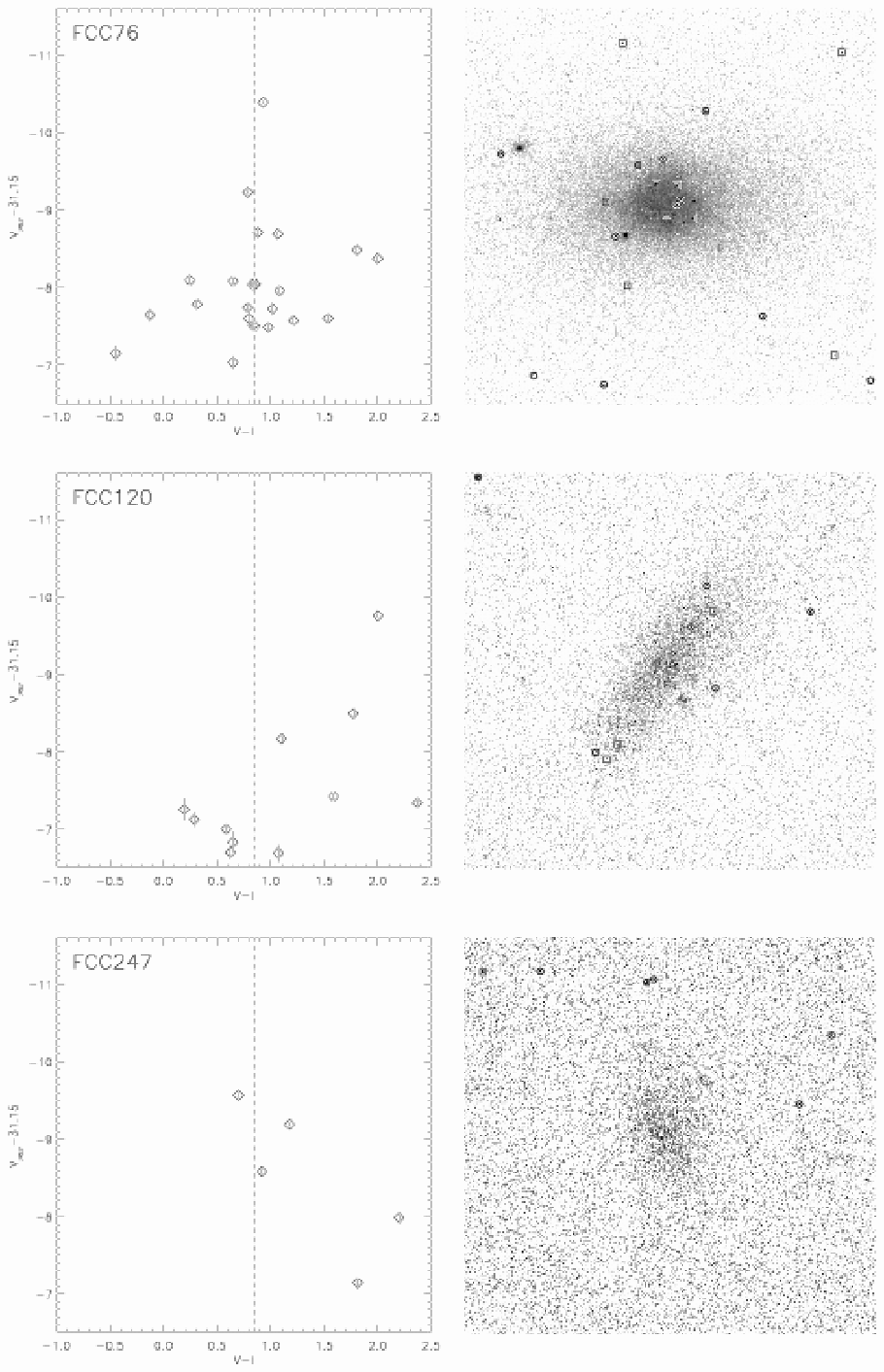}
\end{figure}

\begin{figure}[tp]
\plotone{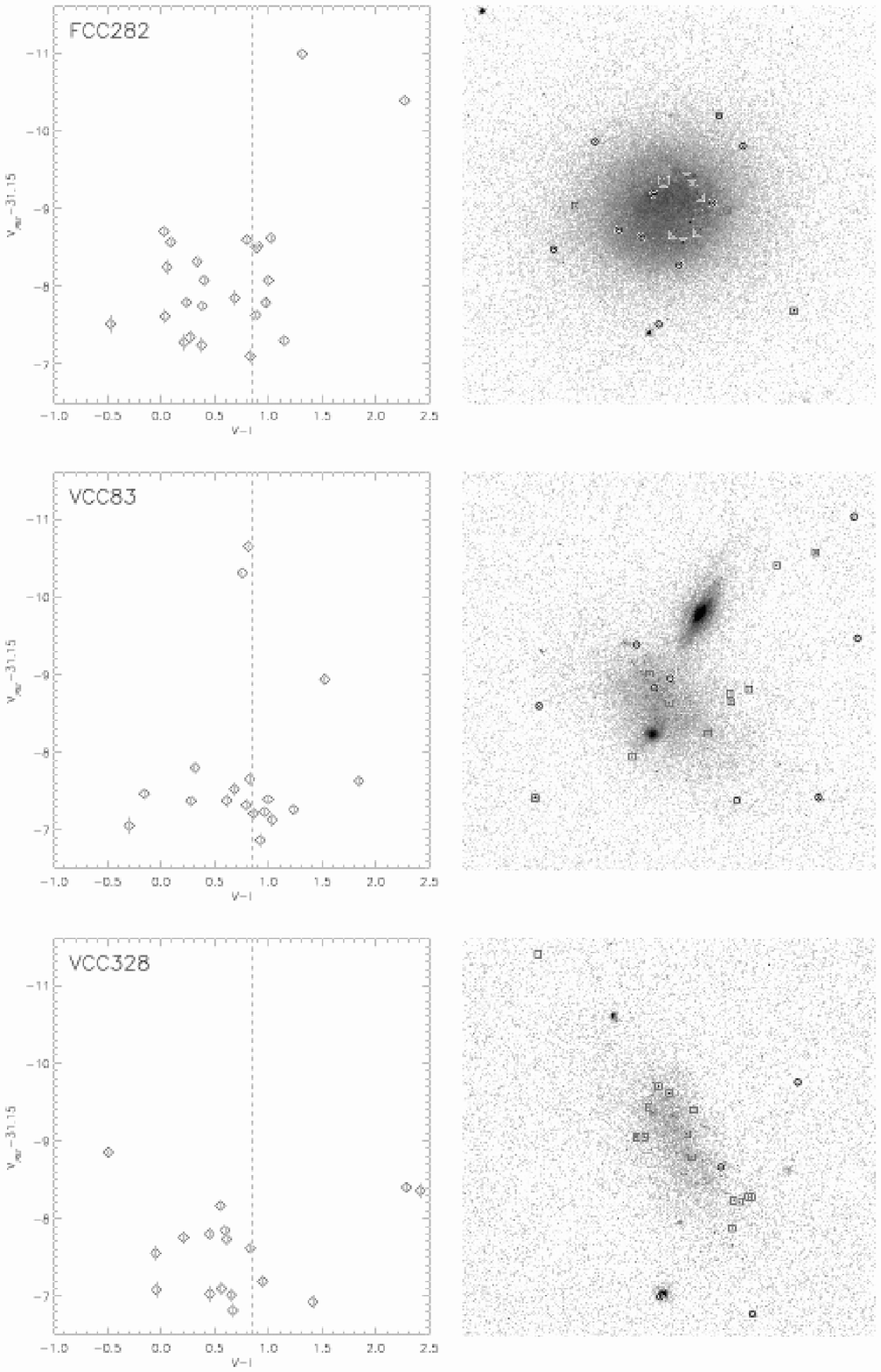}
\end{figure}

\begin{figure}[tp]
\plotone{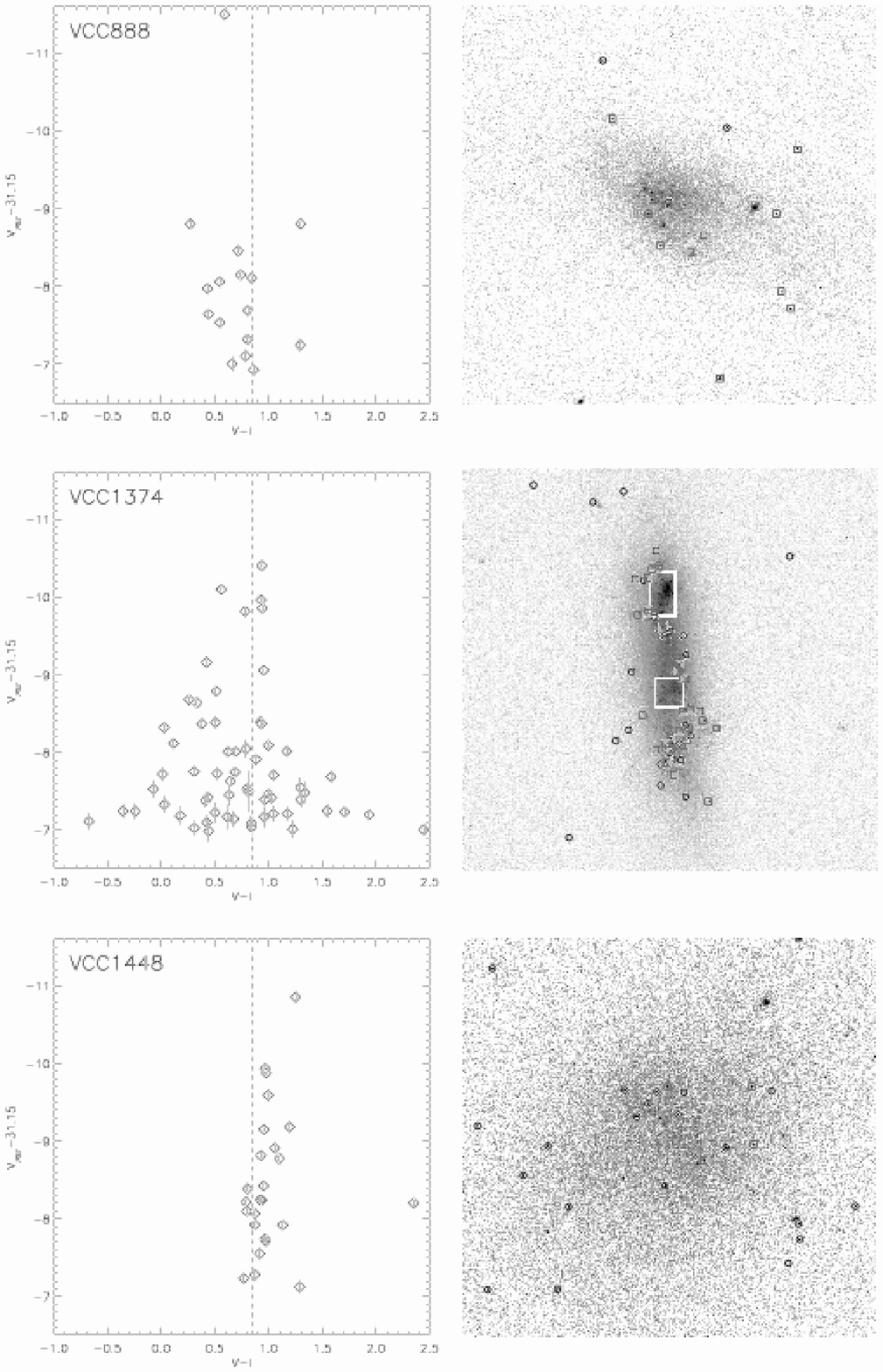}
\end{figure}

\begin{figure}[tp]
\plotone{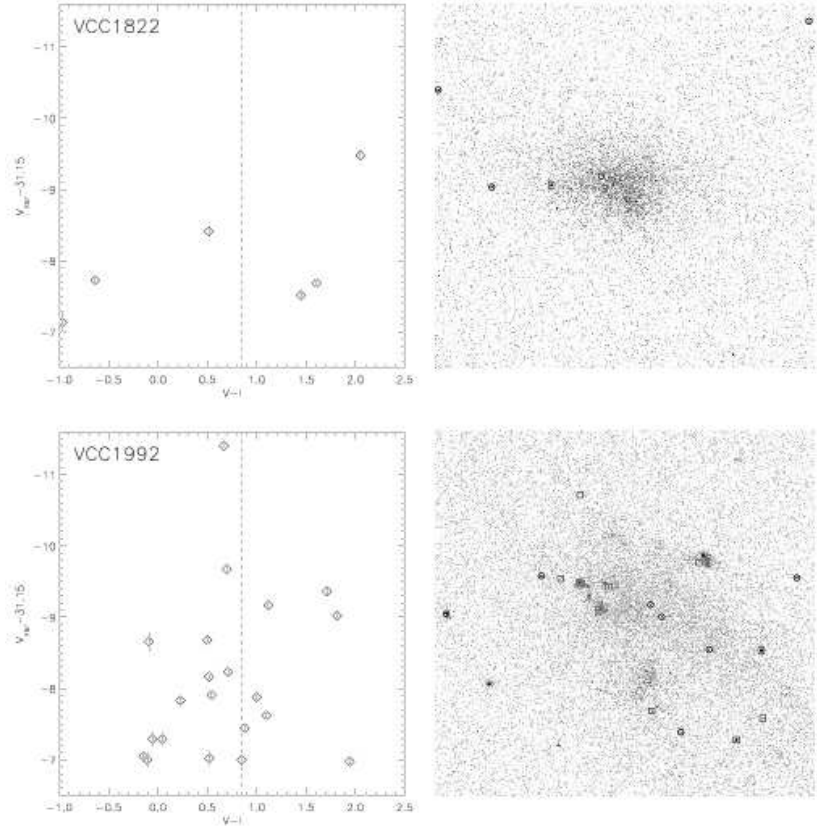}
\caption{{\it Left --} Color magnitude diagram of all cluster
  candidates.  The dashed line at V-I=0.85 represents the dividing
  line between red and blue CCs.  {\it Right --} F555W band images of
  the WF3 chip for each galaxy.  Red cluster candidates are shown as
  circles while blue cluster candidates are denoted with squares.}
\end{figure}

\clearpage

\begin{deluxetable}{lllcrc}
\tablewidth{5in}
\tablecaption{Overview of Observations}
\tablehead{
     \colhead{Galaxy}  &
     \colhead{RA} &
     \colhead{Dec} &
     \colhead{V} &
     \colhead{HST Obs.} &
     \colhead{H$\alpha$ Obs.}  \\
     \colhead{} &
     \colhead{J2000} &
     \colhead{J2000} &
     \colhead{\kms} &
     \colhead{Date} &
     \colhead{}  \\
}

\startdata

FCC 5    &  03:17:51.8 &  -36:45:59 &       &  22/10/98 &      \\
FCC 41 	 &  03:25:35.9 &  -32:57:35 &	    &  16/12/98 &      \\
FCC 76 	 &  03:29:43.2 &  -33:33:25 &  1808\tablenotemark{a} &  12/03/99 &      \\
FCC 120  &  03:33:34.2 &  -36:36:21 &   887\tablenotemark{b} &  14/12/98 &      \\
FCC 128  &  03:34:06.9 &  -36:27:52 &	    &   8/01/99 &      \\
FCC 173  &  03:36:43.2 &  -34:09:33 &	    &  31/07/97 &      \\
FCC 224  &  03:39:32.8 &  -31:47:32 &	    &  31/12/98 &      \\
FCC 247  &  03:40:42.4 &  -35:39:40 &  1097\tablenotemark{a} &  21/02/99 &      \\
FCC 282  &  03:42:45.5 &  -33:55:13 &  1225\tablenotemark{c} &   8/11/98 &      \\
FCC 337  &  03:51:01.2 &  -36:41:00 &	    &  16/12/98 &      \\
VCC 72 	 &  12:13:02.2 &  +14:55:58 &  6351\tablenotemark{d} &  18/12/98 &      \\
VCC 83 	 &  12:13:33.5 &  +14:28:51 &  2441\tablenotemark{d} &  18/12/98 & yes  \\
VCC 280  &  12:18:14.6 &  +11:28:52 &  	    &  16/12/98 &      \\
VCC 328  &  12:19:11.1 &  +12:53:05 &  2179\tablenotemark{d} &   4/01/98 & yes  \\
VCC 364  &  12:19:44.0 &  +12:16:54 &       &  16/12/98 &      \\
VCC 666  &  12:23:46.1 &  +16:47:25 &	    &   2/01/99 &      \\
VCC 888  &  12:26:18.3 &  +08:20:57 &  1096\tablenotemark{d} &  13/12/98 & yes  \\   
VCC 899  &  12:26:27.4 &  +06:42:32 &  4198\tablenotemark{e} &  11/12/98 &      \\
VCC 1165 &  12:29:10.4 &  +09:16:01 &       &  10/12/98 &      \\
VCC 1227 &  12:29:46.8 &  +11:10:02 &	    &  14/12/98 &      \\
VCC 1374 &  12:31:37.7 &  +14:51:38 &  2559\tablenotemark{f} &  19/12/98 & yes  \\
VCC 1413 &  12:32:07.7 &  +12:26:04 &       &  17/12/98 &      \\
VCC 1448 &  12:32:40.7 &  +12:46:16 &  2583\tablenotemark{d} &  13/02/99 & yes  \\ 
VCC 1822 &  12:40:10.4 &  +06:50:48 &  1012\tablenotemark{g} &   7/12/98 & yes  \\
VCC 1889 &  12:41:46.0 &  +11:15:01 &  4725\tablenotemark{e} &  16/12/98 &      \\
VCC 1992 &  12:44:09.7 &  +12:06:47 &  1010\tablenotemark{d} &  21/03/99 & yes  \\
\enddata

\tablecomments{{\footnotesize   
{\it (a) }{\citet{drinkwater01}}
{\it (b) }{\citet{schroder01}}
{\it (c) }{\citet{dacosta98}}
{\it (d) }{\citet{devaucouleurs91}}
{\it (e) }{\citet{binggeli93}}
{\it (f) }{\citet{schneider90}}
{\it (g) }{\citet{binggeli85}}}}

\end{deluxetable}

\begin{deluxetable}{lccccc}
\tablewidth{4.7in}
\tablecaption{Derived Data for all galaxies}
\tablehead{
     \colhead{Galaxy}  &
     \colhead{V$_T$}  &
     \colhead{$\Delta$(V$_T$)}  &
     \colhead{90\% Comp.}  &
     \colhead{\# CC}  &
     \colhead{BG/FG}  \\

     \colhead{} &
     \colhead{} &
     \colhead{} &
     \colhead{R$_{\rm core}$=0} &
     \colhead{} &
     \colhead{} \\
}

\startdata

FCC 5    & 17.5  &  0.09 &  24.38  &   5  &  4.92 $\pm$ 2.59 \\
FCC 41 	 & 18.2  &  1.20 &  24.50  &   3  &  5.27 $\pm$ 2.52 \\ 
FCC 76 	 & 14.6  &  0.04 &  24.38  &  24  &  4.69 $\pm$ 2.53 \\ 
FCC 120  & 15.9  &  0.23 &  24.50  &  12  &  5.04 $\pm$ 2.54 \\ 
FCC 128  & 16.3  &  0.02 &  24.50  &   7  &  5.23 $\pm$ 2.63 \\ 
FCC 173  & 18.7  &  3.79 &  24.45  &   4  &  5.23 $\pm$ 2.70 \\ 
FCC 224  & 16.4  &  0.17 &  24.49  &  17  &  5.30 $\pm$ 2.62 \\ 
FCC 247  & 17.6  &  0.10 &  24.45  &   6  &  5.19 $\pm$ 2.67 \\ 
FCC 282  & 14.2  &  0.01 &  24.36  &  25  &  4.46 $\pm$ 2.52 \\ 
FCC 337  & 17.0  &  0.41 &  24.53  &   4  &  5.42 $\pm$ 2.67 \\ 
VCC 72 	 & 16.2  &  0.07 &  24.30  &   7  &  4.54 $\pm$ 2.50 \\ 
VCC 83 	 & 15.5  &  0.03 &  24.31  &  18  &  4.42 $\pm$ 2.55 \\ 
VCC 280  & 18.4  &  1.58 &  24.29  &   6  &  4.31 $\pm$ 2.36 \\ 
VCC 328  & 16.0  &  0.12 &  24.37  &  18  &  4.54 $\pm$ 2.53 \\ 
VCC 364  & 16.9  &  0.05 &  24.31  &  10  &  4.27 $\pm$ 2.39 \\ 
VCC 666  & 16.6  &  0.06 &  24.36  &   7  &  4.65 $\pm$ 2.51 \\ 
VCC 888  & 15.3  &  0.06 &  24.22  &  16  &  4.19 $\pm$ 2.38 \\ 
VCC 899  & 16.8  &  0.17 &  24.21  &   0  &  4.15 $\pm$ 2.34 \\ 
VCC 1165 & 16.9  &  0.28 &  24.28  &   5  &  4.38 $\pm$ 2.56 \\ 
VCC 1227 & 18.9  &  0.16 &  24.31  &   5  &  4.58 $\pm$ 2.45 \\ 
VCC 1374 & 14.6  &  0.01 &  24.20  &  64  &  3.96 $\pm$ 2.34 \\ 
VCC 1413 & 17.4  &  0.02 &  24.37  &   6  &  4.73 $\pm$ 2.43 \\ 
VCC 1448 & 15.0  &  0.10 &  24.36  &  25  &  4.58 $\pm$ 2.48 \\ 
VCC 1822 & 16.1  &  0.01 &  24.19  &   7  &  3.92 $\pm$ 2.30 \\ 
VCC 1889 & 16.1  &  0.08 &  24.29  &   6  &  4.35 $\pm$ 2.54 \\ 
VCC 1992 & 15.9  &  0.05 &  24.33  &  22  &  4.27 $\pm$ 2.52 \\
\enddata
\end{deluxetable}

\begin{deluxetable}{lccccccc}
\tablecaption{Data for the 11 galaxy subsample}
\tablehead{
     \colhead{Galaxy} &
     \colhead{N$_{\rm blue}$} &
     \colhead{N$_{\rm red}$} &
     \colhead{S$_{\rm N}$} &
     \colhead{Lum. Func.} &
     \colhead{S$_{\rm N}$} &
     \colhead{H$\alpha$} &
     \colhead{SFR} \\

     \colhead{} &
     \colhead{} &
     \colhead{} &
     \colhead{min} &
     \colhead{Correction} &
     \colhead{w/ Corr} &
     \colhead{10$^{38}$ ergs/s} &
     \colhead{M$_\odot$/yr}  \\
}
\startdata
FCC76   &   11 &   13 &      2.3 $\pm$ 0.9 &   1.36  &   3.1 $\pm$ 1.0 & & \\		      
FCC120  &    5 &    7 &      2.4 $\pm$ 2.0 &   1.28  &   3.0 $\pm$ 2.2 & & \\		      
FCC247  &    1 &    5 &      3.7 $\pm$ 7.1 &   1.31  &   4.9 $\pm$ 7.5 & & \\		      
FCC282  &   15 &   10 &      1.1 $\pm$ 0.6 &   1.36  &   1.5 $\pm$ 0.6 & & \\		      
VCC83   &   10 &    8 &      2.5 $\pm$ 1.7 &   1.40  &   3.5 $\pm$ 1.8 &  11.0 &  0.009 \\   
VCC328  &   14 &    4 &      0.5 $\pm$ 1.9 &   1.36  &   0.7 $\pm$ 1.9 &   7.1 &  0.006 \\   
VCC888  &   13 &    3 &     -0.1 $\pm$ 0.9 &   1.47  &  -0.2 $\pm$ 0.9 &   6.3 &  0.005 \\   
VCC1374 &   39 &   25 &      5.3 $\pm$ 1.3 &   1.49  &   7.8 $\pm$ 1.5 &  54.3 &  0.044 \\   
VCC1448 &    4 &   21 &      6.1 $\pm$ 1.7 &   1.36  &   8.3 $\pm$ 1.9 &   0.0 &  0.0 \\     
VCC1822 &    3 &    4 &      1.0 $\pm$ 1.6 &   1.50  &   1.4 $\pm$ 1.7 &   0.2 &  0.0002 \\  
VCC1992 &   15 &    7 &      2.9 $\pm$ 2.1 &   1.39  &   4.0 $\pm$ 2.3 &  16.0 &  0.013 \\   
\enddata
\end{deluxetable}

\end{document}